\documentclass[a4paper,11pt]{article}
\usepackage[top=2.5cm, bottom=2.5cm, left=2cm, right=2cm]{geometry}
\usepackage[utf8]{inputenc}
\usepackage[british]{babel}
\usepackage[mono=false]{libertine}
\usepackage{times}
\usepackage{titlesec}
\usepackage{enumitem}
\usepackage{array}
\usepackage{mathtools}
\usepackage{amssymb,amsfonts,amsmath,amsthm}
\usepackage{dsfont}
\usepackage{bm}
\usepackage{cases}
\usepackage{braket}
\usepackage{bbm}
\usepackage{microtype}
\usepackage[font=small]{caption}
\usepackage{subcaption}
\usepackage[ruled,vlined]{algorithm2e}
\usepackage[perpage,hang,flushmargin]{footmisc} 
\usepackage{titlesec}
\usepackage[usenames, dvipsnames]{color}
\usepackage[titletoc]{appendix}
\usepackage[pdftex]{graphicx}
\usepackage{url}
\usepackage{setspace}
\usepackage{xcolor}
\usepackage[pdfpagemode=UseNone,bookmarksopen=false,colorlinks=true,urlcolor=blue,citecolor=magenta,citebordercolor=blue,linkcolor=blue]{hyperref}

\makeatletter
\newcommand{\setappendix}{Appendix~\thesection:~~}
\newcommand{\setsection}{\thesection~~}
\titleformat{\section}{\bfseries\LARGE}{%
	\ifnum\pdfstrcmp{\@currenvir}{appendices}=0
	\setappendix
	\else
	\setsection
\fi}{0em}{}
\makeatother

\DeclareUnicodeCharacter{00A0}{~}
\newcommand{\bPhi}{\bm{\Phi}}
\newcommand{\bT}{{\textbf {T}}}
\newcommand{\bP}{{\textbf {P}}}
\newcommand{\bO}{{\textbf {O}}}

\newcommand{\bmm}{{\textbf {m}}}

\newcommand{\bU}{{\textbf {U}}}

\newcommand{\bu}{{\textbf {u}}}
\newcommand{\bomega}{{\boldsymbol{\omega}}}
\newcommand{\bV}{{\textbf {V}}}

\newcommand{\bF}{{\textbf {F}}}
\newcommand{\bY}{{\textbf {Y}}}

\newcommand{\bg}{{\textbf {g}}}
\newcommand{\bM}{{\textbf {M}}}
\newcommand{\bbe}{{\textbf {e}}}

\newcommand{\bZ}{{\textbf {Z}}}

\newcommand{\bv}{{\textbf {v}}}

\newcommand{\bX}{{\textbf {X}}}
\newcommand{\bx}{{\textbf {x}}}

\newcommand{\blambda}{{\boldsymbol{\lambda}}}

\newcommand{\bz}{{\textbf {z}}}

\newcommand{\bS}{{\textbf {S}}}

\newcommand{\bR}{{\textbf {R}}}

\newcommand{\bxhat}{{\hat{\textbf {x}}}}
\newcommand{\ee}{\end{equation}}
\newcommand\smallO{
  \mathchoice
    {{\scriptstyle\mathcal{O}}}
    {{\scriptstyle\mathcal{O}}}
    {{\scriptscriptstyle\mathcal{O}}}
    {\scalebox{.7}{$\scriptscriptstyle\mathcal{O}$}}
  }

\newtheorem{theorem}{Theorem}
\newtheorem{conjecture}[theorem]{Conjecture}
\newtheorem{hypothesis}[theorem]{Hypothesis}
\newtheorem{proposition}[theorem]{Proposition}

\newcommand{\EE}{\mathbb{E}}
\newcommand{\bbR}{\mathbb{R}}

\newcommand{\bbK}{\mathbb{K}}

\newcommand{\bbC}{\mathbb{C}}

\DeclareMathOperator*{\extr}{\mathrm{extr}}
\renewcommand{\labelitemi}{$\bullet$}
\allowdisplaybreaks

\begin{document}

\title{Construction of optimal spectral methods in phase retrieval}

\date{}
\author{Antoine Maillard$^{\star,\diamond}$, Florent Krzakala$^{\star,\oplus}$, Yue M. Lu$^\dagger$, Lenka Zdeborov\'a$^{\otimes}$}
\maketitle
{\let\thefootnote\relax\footnote{
$\star$ Laboratoire de Physique de l'\'Ecole Normale Sup\'erieure, PSL University, CNRS, Sorbonne Universit\'es, Paris, France.\\
$\dagger$John A. Paulson School of Engineering and Applied Sciences, Harvard University, Cambridge, MA 02138, USA.\\
$\oplus$ IdePHICS laboratory, EPFL, Switzerland. \\
$\otimes$ SPOC laboratory, EPFL, Switzerland. \\
$\diamond$ To whom correspondence shall be sent: \href{mailto:antoine.maillard@ens.fr}{antoine.maillard@ens.fr}.
}}
\setcounter{footnote}{0}

\begin{abstract}	
        We consider the \emph{phase retrieval} problem, in which the observer wishes to recover a $n$-dimensional real or complex signal $\bX^\star$ from 
        the (possibly noisy) observation
        of $|\bPhi \bX^\star|$, in which $\bPhi$ is a matrix of size $m \times n$. We consider a \emph{high-dimensional} setting where $n,m \to \infty$ with $m/n = \mathcal{O}(1)$, and 
        a large class of (possibly correlated) random matrices $\bPhi$ and observation channels. 
        Spectral methods are a powerful tool to obtain approximate observations of the signal $\bX^\star$ which can be then used as initialization 
        for a subsequent algorithm, at a low 
        computational cost.
        In this paper, we extend and unify previous results and approaches on spectral methods for the phase retrieval problem.
        More precisely, we combine the linearization of message-passing algorithms and the analysis of the \emph{Bethe Hessian}, a classical tool of
        statistical physics.
        Using this toolbox, we show how to derive optimal spectral methods for arbitrary channel noise and right-unitarily invariant matrix $\bPhi$, in an automated manner 
        (i.e.\ with no optimization over any hyperparameter or preprocessing function).
\end{abstract}

\begin{spacing}{1.0}
\tableofcontents
\end{spacing}

\renewcommand{\labelitemi}{$\bullet$}

\section{Introduction}\label{sec:introduction}

\subsection{Setting of the problem and related works}

\noindent
In the phase retrieval problem, one aims to recover an unknown \emph{signal} $\bX^\star \in \bbK^n$ ($\bbK = \bbR$ or $\bbC$) from $m$ measurements $\{y_\mu\}$, which are 
noisy observations of $|\bPhi \bX^\star|$ (the modulus is applied element-wise), with $\bPhi \in \bbK^{m \times n}$ a (random) sensing matrix. 
This model arises in a large set of problems ranging from signal processing \cite{fienup1982phase,unser1988maximum,dremeau2015reference} to statistical estimation \cite{candes2015phase,jaganathan2015phase},
optics, X-ray crystallography, astronomy or microscopy \cite{shechtman2015phase},
where detectors can often only measure information about the amplitude of signals, and lose all information about its phase.
Phase retrieval is also a textbook example of a learning problem with a highly non-convex loss landscape \cite{netrapalli2015phase,sun2018geometric,hand2018phase}.

The majority of algorithms developed to solve this problem are based either on semi-definite programming relaxations \cite{candes2015phasea,waldspurger2018phase,goldstein2018phasemax}
or on more direct non-convex optimization procedures, e.g.\ Wirtinger flow \cite{candes2015phase} or approximate message-passing \cite{schniter2014compressive,mondelli2020approximate} to name a few.
In general, these optimization methods require an ``informed'' initialization $\hat{\bX}$, i.e.\ that is positively correlated with the signal $\bX^\star$. 
The privileged class of algorithms to obtain such initializations in a computationally cheap manner are spectral methods, i.e.\ estimates given by 
the principal eigenvector of an appropriate matrix constructed from the sensing matrix $\bPhi$ and the observations $\{y_\mu\}$ \cite{mondelli2020optimal,luo2019optimal,ma2019spectral}.

In the present work, we consider a \emph{high-dimensional limit} (or thermodynamic limit in the statistical physics language), in which $n,m \to \infty$ with $\alpha \equiv m/n = \Theta(1)$.
In this limit,
a great amount of work is present both in the statistical physics and in the information theory literature for different assumptions 
on the matrix $\bPhi$.
The asymptotic optimal performances in a large class of problems including phase retrieval were conjectured using the non-rigorous
replica method of statistical physics in \cite{kabashima2008inference,takahashi2020macroscopic}, and these results were extended and partly proven in \cite{barbier2019optimal,maillard2020phase}.
Specifically for the phase retrieval problem, the limits of weak-recovery were analyzed for Gaussian matrices $\bPhi$ in \cite{lu2020phase,mondelli2019fundamental,luo2019optimal}, and 
for column-unitary $\bPhi$ in \cite{ma2019spectral,dudeja2020information,dudeja2020analysis}.
In this work we derive the optimal spectral methods for a more generic assumption of right orthogonal (or unitary) invariance, that is we assume:
\begin{hypothesis}[Matrix ensemble]\label{hyp:sensing_matrix}
For every $\bO \in \mathcal{U}(n)$ (or $\mathcal{O}(n)$ in the real case), the following equality holds in distribution $\bPhi \overset{\mathrm{d}}{=} \bPhi \bO$. 
We assume that the spectral measure of $\bPhi^\dagger \bPhi / n$ weakly converges (a.s.)
to a deterministic probability measure $\nu$  and we designate $\langle f(\lambda)\rangle_\nu \equiv \int \nu(\mathrm{d}\lambda) f(\lambda)$ the linear statistics of $\nu$.
\end{hypothesis}
We assume to have access to a factorized prior distribution $P_0$ used to generate $\bX^\star$, with zero mean and variance $\rho\!=\!\EE_{P_0}[|x|^2]\!>\!0$, as well as the ``channel'' distribution $P_\mathrm{out}(y|z)$, giving 
the probability of the observations conditioned on the value of $\bPhi \bX^\star$.
The observations are therefore generated as:
\begin{align}\label{eq:def_Ymu}
    Y_\mu \sim P_\mathrm{out}\Big(\cdot \Big| \frac{1}{\sqrt{n}} \sum_{i=1}^n \Phi_{\mu i} X_i^{\star} \Big), \quad 1 \leq \mu \leq m.
\end{align}
Eq.~\eqref{eq:def_Ymu} defines the very general class of \emph{Generalized Linear Models} (GLMs).
The present work covers a wide class of \emph{phase retrieval} problems, 
in which we assume that $P_\mathrm{out}(y|z)$ is a function of $|z|^2$, and in which the prior distribution $P_0$ is also symmetric: $P_0(x)\!=\!P_0(|x|)$.
The knowledge of $P_0,P_\mathrm{out}$ allows us to discuss the so-called ``Bayes-optimal'' estimator: although somewhat restrictive this knowledge allows for many insightful theoretical studies. 
The information-theoretic and algorithmic limits of 
the models described by eq.~\eqref{eq:def_Ymu} have been rigorously analyzed in \cite{barbier2019optimal,maillard2020phase}.
The Bayes-optimal estimation can be summarized in the study of the \emph{posterior probability} of $\bx$ given the observations $\bY$ and the sensing matrix $\bPhi$:
\begin{align}\label{eq:def_posterior}
    P(\bx | \bY,\bPhi) &\equiv \frac{1}{{\cal Z}_n(\bY)} \prod_{i=1}^n P_0(x_i) \, \prod_{\mu=1}^m P_\mathrm{out}\Big(Y_\mu \Big| \frac{1}{\sqrt{n}} \sum_{i=1}^n \Phi_{\mu i} x_i\Big).
\end{align}
The logarithm of the normalization $(1/n) \ln \mathcal{Z}_n(\bY)$ is usually called the \emph{free entropy} in the statistical physics terminology.
We will generically denote by $\langle \cdot \rangle$ the average with respect to the posterior probability~\eqref{eq:def_posterior} of $\bx$.
A key role in this paper will be played by the algorithmic \emph{weak recovery} threshold, called $\alpha_\mathrm{WR,Algo}$, defined in such a way that for $\alpha < \alpha_\mathrm{WR,Algo}$ all known polynomial-time estimators are uncorrelated with the signal $\bX^\star$, while for $\alpha > \alpha_\mathrm{WR,Algo}$, there exists estimators that recover a finite fraction of the signal in polynomial time. This algorithmic weak recovery threshold depends on the spectral distribution of the matrix $\bPhi$ and of the specific form of the output channel distribution. 
Interestingly, it only depends on the prior distribution $P_0$ via its variance $\rho$.
Its derivation has been presented in \cite{maillard2020phase}, where it was shown that $\alpha_\mathrm{WR,Algo}$ is the only solution to the equation:
\begin{align}\label{eq:wr_threshold_general}
    \alpha = \frac{\langle \lambda \rangle_\nu^2}{\langle \lambda^2 \rangle_\nu}\Big(1+\Big[\int \mathrm{d}y \frac{\Big|\int_\bbK \mathcal{D}_\beta z \ (|z|^2-1) \  P_\mathrm{out}\big(y\big|\sqrt{\frac{\rho \langle \lambda \rangle_\nu}{\alpha}} z\big)\Big|^2}{\int_\bbK \mathcal{D}_\beta z \ P_\mathrm{out}\big(y\big|\sqrt{\frac{\rho \langle \lambda \rangle_\nu}{\alpha}} z\big)}\Big]^{-1}\Big).
\end{align}
In this equation, we let $\beta \in \{1,2\}$, with $\bbK = \bbR$ if $\beta = 1$ and $\bbK = \bbC$ if $\beta = 2$\footnote{The integrals on $\bbC$ are effectively defined as integrals over $\bbR^2$.}.
We introduced the standard Gaussian measure on $\bbK$ as ${\cal D}_\beta z \equiv (2 \pi/\beta)^{-\beta/2} \exp(-\beta |z|^2/2) \mathrm{d}z$.
Note that in eq.~\eqref{eq:wr_threshold_general}, the integrated quantity and the averages $\langle \cdot \rangle_\nu$ depend on $\alpha$, so that this is actually an implicit equation on $\alpha_\mathrm{WR,Algo}$.
An important algorithmic question is to characterize the class of polynomial-time algorithms that 
can achieve weak recovery above the predicted threshold. 

The (generalized) \emph{vector approximate message-passing} (G-VAMP) algorithm \cite{rangan2017vector,schniter2016vector} has been shown to achieve the threshold in \cite{maillard2020phase}.
Furthermore, it has been conjectured to achieve the optimal polynomial-time recovery for rotationally (unitarily) invariant matrices, i.e.\ satisfying Hypothesis~\ref{hyp:sensing_matrix}.
However this algorithm is rather sensitive to the assumptions of the model, that often do not hold in real data:
thus, its applications to real problems are somewhat limited. It is therefore of great interest to investigate more robust and computationally even cheaper algorithms that could achieve similar performances. 
A natural class of such algorithms are spectral methods.
Their output can be used as informative initializations for local gradient-based optimization algorithms, and can induce a jump in the accuracy achieved at a reasonable computational cost. 
Such techniques have already been applied e.g.\ in optical systems \cite{valzania2020accelerating}.
In the context of phase retrieval, the performance of these spectral methods has been rigorously analyzed for Gaussian \cite{lu2020phase,mondelli2019fundamental,luo2019optimal} and unitary \cite{ma2019spectral,dudeja2020analysis,dudeja2020universality} sensing matrices.
For Gaussian sensing matrices, \cite{mondelli2020optimal} also shows how to optimally combine such spectral methods with simple linear estimators, improving even further the performance.

The main goal of the present paper is to design \emph{optimal} spectral methods for the phase retrieval problem in the aforementioned limit, for the very generic class of sensing matrices of Hypothesis~\ref{hyp:sensing_matrix}.
Most importantly, in contrast to the previous aforementioned works our approach is completely \emph{automated}, in the sense the spectral methods we derive are (conjectured to be) optimal without the need for optimization over additional parameters. 
The constructiveness of our approach gives more weight to this optimality conjecture, as we do not restrict to a specific family of spectral methods.

We construct and unify three different approaches for the design of such algorithms, for any sensing matrix satisfying Hypothesis~\ref{hyp:sensing_matrix}: (a) a ``pedestrian'' optimization of the preprocessing function (the approach of the aforementioned previous works), (b) the linearization of message-passing algorithms, and (c)
a \emph{Bethe Hessian} analysis. In short we show that (a) is just a shifted version of (c);
(c) automatically uses the optimal preprocessing function in (a); and two eigenvalues of (b) (the dominant one and a peculiar one) have an exact correspondence with the top eigenvalue of (a).

\subsection{Main results}\label{subsec:main_results}
In most previous approaches \cite{lu2020phase,mondelli2019fundamental,luo2019optimal,ma2019spectral}, the design of spectral methods for the phase retrieval problem was restricted to consider spectra of matrices of the type:
\begin{align}\label{eq:general_class_spectral_methods}
    \bM(\mathcal{T}) &\equiv \frac{1}{n} \sum_{\mu =1}^m \mathcal{T}(y_\mu) \overline{\Phi_{\mu i}} \Phi_{\mu j}.
\end{align}
These matrices are functions of a (bounded) \emph{preprocessing} function $\mathcal{T}$.
It was previously shown for Gaussian i.i.d.\ matrices $\bPhi$ \cite{lu2020phase,luo2019optimal} and for random column-unitary matrices $\bPhi$ \cite{ma2019spectral,dudeja2020analysis} that the optimal transition and reconstruction errors in the class of spectral methods described by eq.~\eqref{eq:general_class_spectral_methods} is attained by the following functions:
\begin{align}\label{eq:Tstar_gaussian_unitary}
    \mathcal{T}^\star_\mathrm{Gaussian}(y) \equiv \frac{\partial_\omega g_\mathrm{out}(y_\mu,0,\rho)}{1 + \rho \partial_\omega g_\mathrm{out}(y_\mu,0,\rho)}, \hspace{1cm}
    \mathcal{T}^\star_\mathrm{Unitary}(y) \equiv 
    \frac{\partial_\omega g_\mathrm{out}(y_\mu,0,\rho/\alpha)}{1 + \frac{\rho}{\alpha}\partial_\omega g_\mathrm{out}(y_\mu,0,\rho/\alpha)}.
\end{align}
In eq.~\eqref{eq:Tstar_gaussian_unitary} we introduced the function $g_\mathrm{out}$, defined as:
\begin{align}\label{eq:def_gout}
  g_\mathrm{out}(y_\mu, \omega, \sigma^2) &\equiv \frac{1}{\sigma^2} \frac{\int_\bbK \mathrm{d}x \ e^{-\frac{\beta}{2 \sigma^2} |x-\omega|^2} \ (x - \omega) \ P_\mathrm{out}(y_\mu | x)}{\int_\bbK \mathrm{d}x \ e^{-\frac{\beta}{2 \sigma^2}|x - \omega|^2}\, P_\mathrm{out}(y_\mu | x)}.
\end{align}
In particular, this implies\footnote{In the complex case, this is the ``Wirtinger'' derivative $\partial_z f(z) \equiv (\partial_x f(z)- i\partial_yf(z))/2$.}:
\begin{align}\label{eq:dwgout_trivial_fixed_point}
  \partial_\omega g_\mathrm{out}(y_\mu, 0, \sigma^2) &= - \frac{1}{\sigma^2} + \frac{1}{\sigma^4}  \frac{\int_\bbK \mathrm{d}x \ e^{-\frac{\beta}{2 \sigma^2} |x|^2} \ |x|^2 \ P_\mathrm{out}(y_\mu | x)}{\int_\bbK \mathrm{d}x \, e^{-\frac{\beta}{2 \sigma^2}|x|^2}\ P_\mathrm{out}(y_\mu | x)}.
\end{align}
Our first result is a conjecture, that generalizes the above two results and gives the optimal spectral method for any phase retrieval problem of the type of eq.~\eqref{eq:def_Ymu} which satisfies Hypothesis~\ref{hyp:sensing_matrix}\footnote{Note that Conjecture~\ref{conj:optimal_spectral_method} is compatible with the results of eq.~\eqref{eq:Tstar_gaussian_unitary}. Indeed, for Gaussian i.i.d.\ matrices, one has $\langle \lambda \rangle_\nu = \alpha$, while for random column-unitary matrices, $\langle \lambda \rangle_\nu = 1$. }:
\begin{conjecture}\label{conj:optimal_spectral_method}
    For any right-rotationally (or unitarily) invariant matrix $\bPhi$ satisfying Hypothesis~\ref{hyp:sensing_matrix}, the optimal (in terms of both weak-recovery transition and achieved reconstruction error) spectral method belongs to the class of eq.~\eqref{eq:general_class_spectral_methods}, and is attained by:
    \begin{align*}
    \mathcal{T}^\star(y) \equiv 
    \frac{\partial_\omega g_\mathrm{out}(y_\mu,0,\rho \langle \lambda \rangle_\nu/\alpha)}{1 + \frac{\rho \langle \lambda \rangle_\nu}{\alpha}\partial_\omega g_\mathrm{out}(y_\mu,0,\rho \langle \lambda \rangle_\nu/\alpha)}.
    \end{align*}
\end{conjecture}
Before detailing further our results, let us explicit two important consequences of Conjecture~\ref{conj:optimal_spectral_method}:
\begin{itemize}[leftmargin=*]
    \item Note that one can always assume the global scaling $\mathrm{Tr}[\bPhi^\dagger \bPhi]/n^2 \to \langle\lambda \rangle_\nu = \alpha$, as it can be absorbed into the channel $P_\mathrm{out}$\footnote{This scaling is chosen to match the one of Gaussian i.i.d.\ matrices.}.
        The optimal spectral method (in terms of weak-recovery threshold and 
        achieved correlation) is then given by $\mathcal{T}^\star(y) = \partial_\omega g_\mathrm{out}(y_\mu,0,\rho)/(1 + \rho\partial_\omega g_\mathrm{out}(y_\mu,0,\rho))$.
        Remarkably, this optimal function
        \emph{does not depend on the spectrum of the sensing matrix $\bPhi$, 
        nor on the sampling ratio $\alpha$}. 
        The universality of the method is striking when one compares the optimal performances achievable both information-theoretically and by message-passing algorithms that are both heavily dependent on the spectrum of the sensing matrix and the sampling ratio $\alpha$, as analyzed in \cite{maillard2020phase}.
        Universality also has deep consequences for phase retrieval practitioners: when using a spectral initialization for a non-convex optimization algorithm, she/he does not have to take into account
        the details of the correlations in $\bPhi$ to construct an optimal spectral method. Although our conjecture requires Hypothesis~\ref{hyp:sensing_matrix}, this assumption can possibly be partially loosened as numerically explored in Section~\ref{sec:numerics}.
        \item Importantly, Conjecture~\ref{conj:optimal_spectral_method} claims optimality of our method among \emph{all spectral methods} that one can construct from the data $\bPhi$ and the observations $\{y_\mu\}$.
        As we will see, it turns out that this optimal method belongs to the class of eq.~\eqref{eq:general_class_spectral_methods}, but our derivation is fully constructive and did not assume anything on the form of the spectral method.
        We believe this is an important improvement of our work with respect to the previous analysis of spectral methods in phase retrieval, which always assumed the method to be in the class of eq.~\eqref{eq:general_class_spectral_methods}.
        In this sense, our work also confirms the validity of this restriction.
\end{itemize}
Our second main result, which is linked to Conjecture~\ref{conj:optimal_spectral_method}, is the reconciliation of different constructions of spectral methods.
In particular, we develop two \emph{automated} approaches to design optimal spectral methods for the phase retrieval problem.
\begin{itemize}[leftmargin=*]
    \item The first approach arises as a linearization of the Generalized \emph{Vector Approximate Message Passing} (G-VAMP) algorithm \cite{schniter2016vector,rangan2017vector}.
    Similar techniques to obtain efficient spectral methods were already investigated in community detection \cite{krzakala2013spectral}, phase retrieval with Gaussian and column-unitary matrices \cite{mondelli2019fundamental,ma2019spectral}, and in the spiked matrix problem \cite{aubin2019spiked} to name a few.
    Here we extend this method to real and complex phase retrieval with a sensing matrix satisfying Hypothesis~\ref{hyp:sensing_matrix}.
    We call $\bM^{\mathrm{(LAMP)}}$ (for Linearized-AMP) the corresponding matrix. It is given by:
    \begin{align}\label{eq:M_LAMP_gout}
    \bM^{(\text{LAMP})} &\equiv \frac{\rho \langle \lambda \rangle_\nu}{\alpha}\Big(\frac{\alpha}{\langle \lambda \rangle_\nu} \frac{\bPhi \bPhi^\dagger}{n} - \mathbbm{1}_m \Big) \mathrm{Diag}(\partial_\omega g_\mathrm{out}(y_\mu, 0, \rho \langle \lambda \rangle_\nu/ \alpha)). 
    \end{align}
    The aforementioned existing works used the \emph{principal eigenvector} $\hat{\bu}$ of this matrix to construct the spectral estimator 
    as $\bxhat_\mathrm{LAMP} \propto \bPhi^\dagger \mathrm{Diag}(\partial_\omega g_\mathrm{out}(y_\mu, 0, \rho \langle \lambda \rangle_\nu/ \alpha)) \hat{\bu}$.
    Interestingly, we will see that this estimator achieves the optimal recovery threshold but \emph{sub-optimal performance}.
    In Section~\ref{subsec:unification}, we show that the optimal estimator can also be derived from the spectrum of $\bM^{\mathrm{(LAMP)}}$ but that it 
     is ``hidden'' inside the bulk of $\bM^{\mathrm{(LAMP)}}$.
    \item Our second approach leverages the Thouless-Anderson-Palmer (TAP) formalism of statistical physics \cite{thouless1977solution}, using the results of \cite{maillard2019high}. 
    The TAP approach consists in studying the posterior distribution of eq.~\eqref{eq:def_posterior} by ``tilting'' it in a controllable manner: this allows to study a modified posterior distribution 
    in which the first and second moments of all $x_i$ are fixed. These moments become then variables of the free energy associated with this modified posterior distribution:
    this is called the \emph{TAP free energy} in the statistical physics language.
    When weak recovery of the signal is impossible, this free energy possesses a global minimum in the completely uninformative point in which 
    the estimator is the vector $\bmm = 0$. 
    On the other hand, when weak recovery is possible, the optimal estimator corresponds to the global minimum of the TAP free energy with $\bmm \neq 0$. 
    However ws we will see the point $\bmm = 0$ always remains a stationary point of the TAP free energy.
    Moreover, a spectral method used for initializing a non-convex optimization algorithm can be based solely on the observations (i.e.\ on $\bPhi$ and $\{y_\mu\}$), and therefore can not exploit any physical information other than the one present in the uninformative point. 
    When this point is locally stable, we therefore expect all polynomial-time algorithms not to be able to achieve weak recovery. 
    This conjecture has been proven in some cases, e.g.\ in \cite{mondelli2020optimal} for Gaussian $\bPhi$, in \cite{dudeja2020information} for unitary $\bPhi$, and in \cite{maillard2020phase} for a large class of 
    right-rotationally invariant $\bPhi$.
    On the other hand, linear instability of the $\bmm = 0$ point implies that there should exist a minimum of the TAP free entropy with positive correlation with the signal, and which corresponds to the optimal estimator.
    With this picture in mind, it is natural to conjecture that the optimal spectral estimator is the dominant unstable direction of the uninformative fixed point, i.e.\ the smallest eigenvalue of the Hessian.
    Indeed, this is the most informative direction that one can obtain solely by a local analysis of the $\bmm = 0$ point.
    The Hessian of the TAP free energy at the uninformative point is also denoted \emph{Bethe Hessian}.
    Notably, this Bethe Hessian has been investigated in the context of community detection \cite{saade2014spectral}.
    This leads to another method, called $\bM^{\mathrm{(TAP)}}$, which is (up to a shift) the method $\bM(\mathcal{T}^\star)$ given in Conjecture~\ref{conj:optimal_spectral_method}:
    \begin{align}\label{eq:def_MTAP}
        \bM^{\mathrm{(TAP)}} &\equiv - \frac{1}{\rho} \mathbbm{1}_n + \frac{1}{n} \sum_{\mu = 1}^m \frac{\partial_\omega g_\mathrm{out}(y_\mu,0,\rho \langle \lambda \rangle_\nu/\alpha)}{1 + \frac{\rho \langle \lambda \rangle_\nu}{\alpha}\partial_\omega g_\mathrm{out}(y_\mu,0,\rho \langle \lambda \rangle_\nu/\alpha)} \overline{\Phi_{\mu i}} \Phi_{\mu j}.
    \end{align}
\end{itemize}
Let us now briefly outline the structure of the paper.
In Section~\ref{sec:spectral}, we unify three different approaches to construct optimal spectral methods for the phase retrieval problem. 
The first one, based on linearizing the vector approximate message passing is studied in Section~\ref{subsec:lamp},
In Section~\ref{subsec:tap} we consider a second approach, based on the Bethe Hessian.
Remarkably, as we show in Section~\ref{subsec:unification}, for any channel distribution and sensing matrix $\bPhi$, this method coincides exactly with the third approach, which consists in simply generalizing a spectral method that has been proven 
to be optimal for Gaussian \cite{luo2019optimal} and unitary \cite{dudeja2020information} sensing matrices, see eq.~\eqref{eq:Tstar_gaussian_unitary}.
We relate the performance of these different approaches, and show that they allow to conjecture the optimal spectral method,
summarized in Conjecture~\ref{conj:optimal_spectral_method}.
In Section~\ref{sec:numerics}, we give numerical evidence to support our claims.
We give the performance of the spectral methods we derived in phase retrieval, for noiseless and Poisson-noisy observations. 
We also show that our methods perform very well even by allowing more structure in the sensing matrix than assumed in Hypothesis~\ref{hyp:sensing_matrix}, by considering for example randomly subsampled DFT, Hadamard or DCT matrices\footnote{Note that the universality of linearized approximate message passing algorithms for a Gaussian prior and different 
ensembles of column-orthogonal matrices was analyzed recently in \cite{dudeja2020universality}.}.

\textbf{Notations -}
Before presenting the technical aspects of our work, we introduce some notations.
Recall that $\beta = 1,2$ for respectively real and complex variables.
$\mathcal{U}_\beta(n)$ denotes the orthogonal (or unitary) group.
For $x,y \in \bbK$, we define
$x \cdot y \equiv xy$ if $\bbK = \bbR$ and $x \cdot y \equiv \mathrm{Re}[\overline{x}y]$ if $\bbK = \bbC$. 
\section{Spectral methods, message-passing algorithms and TAP approach}\label{sec:spectral}

\subsection{Linearized vector approximate message passing}\label{subsec:lamp}

In this section, we describe the vector approximate message-passing algorithm for the phase retrieval  problem with sensing matrices satisfying Hypothesis~\ref{hyp:sensing_matrix}. 
The algorithm was first stated in \cite{rangan2017vector} for the compressed sensing problem, and later generalized in \cite{schniter2016vector} 
to any GLM described by eq.~\eqref{eq:def_Ymu}.
It makes use of the SVD decomposition of $\bPhi$, that we write as $\bPhi/\sqrt{n} = \bU \bS \bV^\dagger$.
The full iterations of the algorithm are detailed in Algorithm~\ref{algo:gvamp}.
\begin{algorithm}[t]
\caption{Generalized Vector Approximate Message Passing \label{algo:gvamp}}
  \KwData{The sensing matrix $\bPhi/ \sqrt{n} = \bU \bS \bV^\dagger$, the outputs $\{y_\mu\}_{\mu = 1}^m$, a number of iterations $T$.}
  \KwResult{An estimate $\bxhat$ of $\bX^\star$.}
  Randomly initialize all variables\; 
  \For{$t=1,\cdots,T$}{
    \begin{tabular}{ m{0.4\textwidth} m{0.4\textwidth}} 
  \emph{(Denoising $\bx$)}  & \emph{(Denoising $\bz \equiv \frac{1}{\sqrt{n}}\bPhi \bx$})  \\
  $\hat{\bx}_1^t = g_{x1}(\bT_{1}^t, \gamma_{1}^t)$   &  $\hat{\bz}_1^t = g_{z1}(\bR_{1}^t, \tau_{1}^t)$\\
  $v_{1}^t = \frac{1}{\beta n} \sum_{i=1}^n \partial_{T_i} g_{x1}(\bT_{1}^t, \gamma_{1}^t)$  & $c_{1}^t = \frac{1}{\beta m} \sum_{\mu=1}^m \partial_{R_\mu} g_{z1}(\bR_{1}^t, \tau_{1}^t)$ \\
  $\bT_{2}^t = \frac{1}{v_1^t} \hat{\bx}_1^t - \bT_1^t$ & $\bR_{2}^t = \frac{1}{c_1^t} \hat{z}_1^t - \bR_1^t$\\
  $\gamma_{2}^t = \frac{1}{v_1^t} - \gamma_1^t $ & $\tau_{2}^t = \frac{1}{c_1^t} - \tau_1^t $ \\
\end{tabular} \\
 \begin{tabular}{ m{0.4\textwidth} m{0.4\textwidth}} 
  \emph{(Estimation of $\bx$)} & \emph{(Estimation of $\bz$)} \\
  $\hat{\bx}_2^t = g_{x2}(\bT_{2}^t,\bR_2^t,\gamma_2^t,\tau_2^t)$  & $\hat{\bz}_2^t = g_{z2}(\bT_{2}^t,\bR_2^t,\gamma_2^t,\tau_2^t)$ \\
  $v_2^t =  \Big \langle\frac{1}{\tau_2^t \lambda + \gamma_2^t}\Big\rangle_\nu$ & $c_2^t = \frac{1}{\alpha} \Big\langle\frac{\lambda}{\tau_2^t \lambda + \gamma_2^t}\Big\rangle_\nu$ \\
  $\bT_{1}^{t+1} = \frac{1}{v_2^t} \hat{\bx}_2^t - \bT_2^t$& $\bR_{1}^{t+1} = \frac{1}{c_2^t} \hat{\bz}_2^t - \bR_2^t$\\
  $\gamma_1^{t+1} = \frac{1}{v_2^t} - \gamma_2^t$  & $\tau_1^{t+1} = \frac{1}{c_2^t} - \tau_2^t$ \\
  \end{tabular}
  }
  \Return{$\bxhat_1^T$} \;
  \smallskip
\end{algorithm}
We used some auxiliary functions, defined below:
\begin{align}
  \label{eq:auxiliary_functions_gvamp}
  \hspace{-0.3cm}
  \begin{cases}
  g_{x1}(\bT,\gamma)_i \equiv \EE_{P_0(\gamma,-T_i)}[x], &\hspace{-0.1cm}g_{x2}(\bT,\bR,\gamma,\tau) \equiv \frac{\bT}{\gamma} + \bV \bS^\intercal \big(\frac{\gamma}{\tau} + \bS \bS^\intercal\big)^{-1} \big(\frac{\bU^\dagger \bR}{\tau} - \frac{\bS \bV^\dagger \bT}{\gamma}\big), \\
  g_{z1}(\bR,\tau)_\mu \equiv \EE_{P_\mathrm{out}\big(y_\mu,\frac{R_\mu}{\tau} ,\frac{1}{\tau}\big)}[z], &\hspace{-0.1cm} g_{z2}(\bT,\bR,\gamma,\tau) \equiv \bU \bS \bV^\dagger g_{x2}(\bT,\bR,\gamma,\tau).
  \end{cases}
\end{align}
We denoted $P_0(\gamma,\lambda)$ the probability distribution with density proportional to $P_0(x) e^{-\frac{\beta \gamma}{2} |x|^2- \beta \lambda_i \cdot x}$, and $P_\mathrm{out}(y_\mu,\omega_\mu,b)$ the one with density proportional to $P_\mathrm{out}(y_\mu|z) e^{-\frac{\beta|z-\omega_\mu|^2}{2 b}}$.

\subsubsection{The trivial fixed point}

In Algorithm~\ref{algo:gvamp}, one can use the Bayes-optimality hypothesis to derive the following relation (see for instance eq.~(107) of \cite{kabashima2016phase}):
\begin{align}
  \label{eq:bayes_optimal_gout}
 \frac{1}{m} \sum_{\mu=1}^m \EE_{P_\mathrm{out}\big(y_\mu,\frac{(R_1^t)_\mu}{\tau_1^t} ,\frac{1}{\tau_1^t}\big)}\Big[\big|z - \frac{(R_1^t)_\mu}{\tau_1^t}\big|^2\Big] &= \frac{1}{\tau_1^t}.
\end{align}
Informally, eq.~\eqref{eq:bayes_optimal_gout} expresses that the estimated variance of $\bz$, defined as $\tau_1^t$, is equal to the mean square difference between $\bz$ and the estimation of $\bz$ (being $\bR_1^t / \tau_1^t$) under the estimated posterior. 
Recall that we assumed that $P_0$ is symmetric with $\rho \equiv \EE_{P_0}[|x|^2]$ and that $P_\mathrm{out}(y|z)$ only depends on $|z|^2$.
Using eq.~\eqref{eq:bayes_optimal_gout} along with this hypothesis, it is easy to see that Algorithm~\ref{algo:gvamp}
admits the following fixed point, that we call ``trivial'' as it is completely uninformative:
\begin{align}\label{eq:trivial_fixed_point_gvamp}
\left\lbrace\,
\begin{array}{@{}l@{\quad}l@{\quad}l@{\quad}l@{}}
    \gamma_1 = 0, &\gamma_2 = \rho^{-1}, &v_1 = \rho, &v_2 = \rho \\
    \hat{\bx}_1 = \bT_1 = 0, &\hat{\bx}_2 = \bT_2 = 0, &\tau_1 = \alpha / (\rho \langle \lambda \rangle_\nu), & \tau_2 = 0 \\
    c_1 = \rho \langle \lambda \rangle_\nu / \alpha, &c_2 =  \rho \langle \lambda \rangle_\nu / \alpha, &\hat{\bz}_1 = \bR_1 = 0, & \hat{\bz}_2 = \bR_2 = 0.
\end{array}
\right.
\end{align}

\subsubsection{Linearization around the fixed point}

We can now linearize Algorithm~\ref{algo:gvamp} around the fixed point given by eq.~\eqref{eq:trivial_fixed_point_gvamp}.
We begin by showing that the first order variations of all the variances and inverse variances parameters are negligible, 
and we detail this derivation in Appendix~\ref{subsec:app_variances_lamp}.
This will greatly simplify our linearization around the trivial fixed point, as we can 
focus solely on the vector parameters. 
For clarity, we restrict here to the real case $\beta = 1$, while the derivation in the complex case 
is provided in Appendix~\ref{sec_app:lamp_derivation}.
We write the linearization of Algorithm~\ref{algo:gvamp} as (all derivatives are taken at the fixed point of eq.~\eqref{eq:trivial_fixed_point_gvamp}):
\begin{align}\label{eq:linearization_trivial_fixed_point_vector}
  \hspace{-0.3cm}
\begin{array}{@{}l@{\quad}l@{\quad}l@{}}
    \delta \hat{\bx}_1^t = \nabla_{\bT} g_{x1} \delta \bT_1^t, &\delta \hat{\bz}_1^t = \nabla_{\bR} g_{z1} \delta \bR_1^t, &\delta \bT_2^t = \frac{1}{\rho} \delta \hat{\bx}_1^t - \delta \bT_1^t, \\
    \delta \hat{\bx}_2^t = \nabla_{\bT} g_{x2} \delta \bT_2^t  + \nabla_{\bR} g_{x2} \delta \bR_2^t, &  &\delta \hat{\bz}_2^t = \nabla_{\bT} g_{z2} \delta \bT_2^t  + \nabla_{\bR} g_{z2} \delta \bR_2^t, \\
    \delta \bR_2^t = \frac{\alpha}{\rho \langle \lambda \rangle_\nu} \delta \hat{\bz}_1^t - \delta \bR_1^t, &\delta \bT_1^{t+1} = \frac{1}{\rho} \delta \hat{\bx}_2^t - \delta \bT_2^t, & \delta \bR_1^{t+1} = \frac{\alpha}{\rho \langle \lambda \rangle_\nu} \delta \hat{\bz}_2^t - \delta \bR_2^t. 
\end{array}
\end{align}
The derivatives of the auxiliary functions of eq.~\eqref{eq:auxiliary_functions_gvamp} at the trivial fixed point of eq.~\eqref{eq:trivial_fixed_point_gvamp} are:
\begin{align}\label{eq:auxiliary_functions_trivial_fixed_point}
\left\lbrace\,
\begin{array}{@{}l@{\quad}l@{\quad}l@{}}
  \partial_{T_j} [(g_{x1})_i] = \rho \, \delta_{ij}, &\partial_{T_j} [(g_{x2})_i] = \rho \, \delta_{ij}, &\partial_{R_\nu} [(g_{z1})_\mu] = \delta_{\mu \nu} \EE_{P_\mathrm{out}\big(y_\mu,0 ,\frac{\rho \langle \lambda \rangle_\nu}{\alpha}\big)}[z^2], \\
  \partial_{R_\mu} [(g_{x2})_i] = \rho \frac{(\bPhi^\dagger)_{i \mu}}{\sqrt{n}}, &\partial_{T_i} [(g_{z2})_\mu] = \rho \, \frac{\Phi_{\mu i}}{\sqrt{n}}, &\partial_{R_\nu} [(g_{z2})_\mu] = \rho \, \frac{(\bPhi \bPhi^\dagger)_{\mu \nu}}{n}.
\end{array}
\right.
\end{align}
Plugging eq.~\eqref{eq:auxiliary_functions_trivial_fixed_point} in eq.~\eqref{eq:linearization_trivial_fixed_point_vector} yields, with $v(y_\mu) \equiv \EE_{P_\mathrm{out}\big(y_\mu,0 ,\frac{\rho \langle \lambda \rangle_\nu}{\alpha}\big)}[z^2]$:
\begin{align}\label{eq:linearization_trivial_fixed_point_vector_2}
\left\lbrace\,
\begin{array}{@{}l@{\quad}l@{\quad}l@{}}
    \delta \hat{\bx}_1^t = \rho \delta \bT_1^t, & \delta \hat{\bz}_1^t = \mathrm{Diag}(\{v(y_\mu)\}) \delta \bR_1^t, &\delta \bT_2^t = \frac{1}{\rho} \delta \hat{\bx}_1^t - \delta \bT_1^t, \\
    \delta \bR_2^t = \frac{\alpha}{\rho \langle \lambda \rangle_\nu} \delta \hat{\bz}_1^t - \delta \bR_1^t,& \delta \hat{\bx}_2^t = \rho \delta \bT_2^t  + \rho \frac{\bPhi^\dagger}{\sqrt{n}} \delta \bR_2^t, & \delta \hat{\bz}_2^t = \rho \frac{\bPhi}{\sqrt{n}} \delta \bT_2^t  + \rho \frac{\bPhi \bPhi^\dagger}{n} \delta \bR_2^t, \\
    \delta \bT_1^{t+1} = \frac{1}{\rho} \delta \hat{\bx}_2^t - \delta \bT_2^t, & \delta \bR_1^{t+1} = \frac{\alpha}{\rho \langle \lambda \rangle_\nu} \delta \hat{\bz}_2^t - \delta \bR_2^t. &
\end{array}
\right.
\end{align}
These equations imply $\delta \bT_2^t = 0$. The equations can then simply be closed on $\delta \bR_1^t$:
\begin{align}\label{eq:linearization_gvamp_final}
    \delta \bR_1^{t+1} &= \Big(\frac{\alpha}{\langle \lambda \rangle_\nu} \frac{\bPhi \bPhi^\dagger}{n} - \mathbbm{1}_m \Big) \Big[\frac{\alpha}{\rho \langle \lambda \rangle_\nu} \mathrm{Diag}(\{v(y_\mu)\}) - \mathbbm{1}_m\Big] \delta \bR_1^t.
\end{align}
As shown in Appendix~\ref{sec_app:lamp_derivation}, we obtain the same equation in the complex case.
Interestingly, $v(y_\mu)$ can be linked to the function $\partial_\omega g_\mathrm{out}$, simply by eq.~\eqref{eq:dwgout_trivial_fixed_point}:
$\partial_\omega g_\mathrm{out}(y_\mu, 0, \sigma^2) = -\sigma^{-2} + \sigma^{-4} v(y_\mu)$.

\subsubsection{The LAMP spectral method}

The Linearized-AMP (LAMP) spectral method is based on eq.~\eqref{eq:linearization_gvamp_final}, and consists in taking the largest eigenvalue and corresponding eigenvector of the $m \times m$ matrix:
\begin{align*}
  \bM^{(\text{LAMP})} &\equiv \frac{\rho \langle \lambda \rangle_\nu}{\alpha}\Big(\frac{\alpha}{\langle \lambda \rangle_\nu} \frac{\bPhi \bPhi^\dagger}{n} - \mathbbm{1}_m \Big) \mathrm{Diag}(\partial_\omega g_\mathrm{out}(y_\mu, 0, \rho \langle \lambda \rangle_\nu/ \alpha)).
\end{align*}
Note that $\bM^{(\text{LAMP})}$ is not a Hermitian matrix, so ``largest'' eigenvalue means here eigenvalue of largest real part.
If $\hat{\bu}$ is the eigenvector of $\bM^{(\text{LAMP})}$ associated to this largest eigenvalue, then 
one can construct a corresponding estimate $\hat{\bx}$ using the relations of eq.~\eqref{eq:linearization_trivial_fixed_point_vector_2}, as:
\begin{align}\label{eq:correspondance_LAMP_TAP_estimators}
  \hat{\bx} &\equiv \frac{ \bPhi^\dagger \big[\frac{\alpha}{\rho \langle \lambda \rangle_\nu} \mathrm{Diag}(\{v(y_\mu)\}) - \mathbbm{1}_m\big] \hat{\bu}}{\Big\lVert \bPhi^\dagger\big[\frac{\alpha}{\rho \langle \lambda \rangle_\nu} \mathrm{Diag}(\{v(y_\mu)\}) - \mathbbm{1}_m\big] \hat{\bu} \Big \rVert} \sqrt{n \rho}.
\end{align}
Surprisingly, and as we will see in more details in Sections~\ref{subsec:unification} and \ref{sec:numerics}, this spectral method achieves 
the optimal weak-recovery threshold but only sub-optimal performance compared to $\bM(\mathcal{T}^\star)$.
There is, however, a way to recover the optimal performance from $\bM^{(\text{LAMP})}$ by considering an eigenvalue equal to $1$ (and thus ``hidden'' inside the bulk)
which appears when weak recovery is possible.

\subsection{The Bethe Hessian approach}\label{subsec:tap}

\subsubsection{The TAP free entropy}\label{subsubsec:tap_free_entropy}
In this section we detail a statistical-physics based constructive approach to derive the optimal spectral method for the phase retrieval problem.
We consider the so-called \emph{Thouless-Anderson-Palmer} (TAP) \cite{thouless1977solution} free entropy of the system, that we denote $f_\mathrm{TAP}(\bY,\bPhi,\bmm,\sigma)$. The idea of this approach is to constrain the posterior probability of eq.~\eqref{eq:def_posterior} to satisfy the first and second moment constraints $\langle \bx \rangle = \bmm$, $\langle \lVert\bx - \langle \bx \rangle\rVert^2 \rangle = n \sigma^2$, and to study the free entropy of this ``tilted'' probability distribution. 
This provides a dual perspective on the posterior distribution (also called Gibbs measure), by considering the
landscape of $f_\mathrm{TAP}$.
For clarity we will drop the dependency of $f_\mathrm{TAP}$ on $\bY,\bPhi$.
Of particular interest are the maxima of this free entropy, corresponding to \emph{pure states} in the statistical physics language.
It is known that the fixed points of the optimal algorithm for this problem, i.e.\ generalized \emph{vector approximate message-passing} (see Section~\ref{subsec:lamp}),
are in exact correspondence with the local maxima of the TAP free entropy. 
This is shown in \cite{maillard2019high}, in which the TAP free entropy for rotationally-invariant generalized linear models is also derived\footnote{The results of \cite{maillard2019high} stand in the real case, but can be straightforwardly generalized to complex variables.}.
By maximizing as well on the variance parameter $\sigma^2$, it yields, up to $\smallO_n(1)$ terms:
\begin{align}\label{eq:TAP_free_entropy}
   f_\mathrm{TAP}&(\bmm) = \sup_{\sigma \geq 0} \sup_{\substack{\bg \in \bbK^m \\ r \geq 0}} \extr_{\substack{\bomega \in \bbK^m \\ b \geq 0}} \extr_{\substack{\blambda \in \bbK^n \\ \gamma \geq 0}} \Big[\frac{\beta}{n} \sum_{i=1}^n \lambda_i \cdot m_i  + \frac{\beta \gamma}{2n}\big(n \sigma^2 + \sum_{i=1}^n |m_i|^2\big) - \frac{\beta}{n} \sum_{\mu=1}^m \omega_\mu \cdot g_\mu \nonumber \\ 
   &-\frac{\beta b}{2 n} \big(\sum_{\mu=1}^m |g_\mu|^2 -\alpha n r\big)  + \frac{1}{n} \sum_{i=1}^n \ln \int_\bbK P_0(\mathrm{d}x) e^{-\frac{\beta \gamma}{2} |x|^2- \beta \lambda_i \cdot x}  \\ 
   &+ \frac{\alpha}{m} \sum_{\mu=1}^m \ln \int_\bbK \frac{\mathrm{d}h}{\big(\frac{2 \pi b}{\beta}\big)^{\beta/2}} P_\mathrm{out}(y_\mu|h) e^{-\frac{\beta|h-\omega_\mu|^2}{2 b}} + \frac{\beta}{n} \sum_{i=1}^n \sum_{\mu=1}^m g_\mu \cdot \big(\frac{\Phi_{\mu i}}{\sqrt{n}}  m_i\big) + \beta F(\sigma^2,r)\Big]. \nonumber
\end{align}
Here the notation $\extr_\gamma f(\gamma)$ means that one should solve the corresponding saddle-point equation $\partial_\gamma f(\gamma) = 0$, and the function $F$ is defined as:
\begin{align*}
    F(x,y) &\equiv \inf_{\zeta_x,\zeta_y > 0} \Big[\frac{\zeta_x x}{2} + \frac{\alpha \zeta_y y}{2} - \frac{\alpha-1}{2}\ln \zeta_y - \frac{1}{2} \langle \ln(\zeta_x \zeta_y + \lambda)\rangle_\nu\Big] - \frac{1}{2} \ln x - \frac{\alpha}{2} \ln y - \frac{1+\alpha}{2}.
\end{align*}
One can write the saddle-point equations associated to eq.~\eqref{eq:TAP_free_entropy}, called the \emph{TAP equations}:
\begin{align}\label{eq:tap_eqs_complete}
\begin{array}{@{}l@{\quad}l@{\quad}l@{}}
     m_i = \EE_{P_0(\gamma,\lambda_i)}[x], & & \hspace{-3.3cm} \sigma^2 = \frac{1}{n} \sum\limits_{i=1}^n \EE_{P_0(\gamma,\lambda_i)}[|x - m_i|^2], \\ 
     g_\mu = g_\mathrm{out}(y_\mu,\omega_\mu,b), & & \hspace{-3.3cm}r = \frac{1}{m} \sum\limits_{\mu=1}^m \Big\{ |g_\mu|^2 + \frac{1}{b} - \EE_{P_\mathrm{out}(y_\mu,\omega_\mu,b)}\Big[\big|\frac{h-\omega_\mu}{b}\big|^2\Big] \Big\}, \\ 
     \omega_\mu + b g_\mu = \sum\limits_{i=1}^n \frac{\Phi_{\mu i}}{\sqrt{n}} m_i, & b  = - \frac{2}{\alpha} \partial_r F(\sigma^2,r), & \gamma = - 2 \partial_{\sigma^2} F(\sigma^2,r).
\end{array}
\end{align}

\subsubsection{The trivial fixed point}

It is easy to see that the TAP equations~\eqref{eq:tap_eqs_complete} admits a trivial fixed point at $\bmm = 0$ (corresponding to a local maximum of $f_\mathrm{TAP}$).
At this point, the parameters are $\sigma^2 = \rho$, $\bg = \bomega = \blambda = 0$, $\gamma = r = 0$, $b = \rho \langle \lambda \rangle_\nu / \alpha$. 
This uses in particular a known consequence of the Bayes-optimality, that relates the variance parameter $b$
to the mean squared difference between the true $\bPhi \bX^\star$ and its estimate, see \cite{kabashima2016phase}\footnote{This relation is equivalent to eq.~\eqref{eq:bayes_optimal_gout}, which states it for AMP iterations rather than the solutions of the TAP equations.}:
$\frac{1}{m} \sum_{\mu=1}^m  \EE_{P_\mathrm{out}(y_\mu,\omega_\mu,b)}[|h-\omega_\mu|^2] = b$.
The derivation of the fixed point also uses the behavior of $F(\sigma^2,r)$ at small $r$, computed in Appendix~\ref{subsec:app_expansion_F}:
\begin{align}\label{eq:F_small_r}
F(\sigma^2,r) = -\frac{\langle \lambda \rangle_\nu r \sigma^2}{2} + \frac{\sigma^4 r^2}{4 \alpha} [\alpha \langle \lambda^2\rangle_\nu - (1+\alpha) \langle \lambda \rangle_\nu^2] + \sigma^6 r^3 G(r \sigma^2),
\end{align}
with $G(x)$ a continuous bounded function in $x = 0$.

\subsubsection{The spectral method}

A natural way to design a spectral method for this inference problem is to consider the Hessian of $-f_\mathrm{TAP}$ at this trivial fixed point, as we expect a descending informative direction to appear in its spectrum at the weak recovery threshold. 
As we show in Appendix~\ref{sec:app_hessian_tap}, this procedure leads to consider the $n \times n$ matrix:
\begin{align*}
    \bM^{\mathrm{(TAP)}} &\equiv - \frac{n}{\beta}\nabla^2 f_\mathrm{TAP}(0) = - \frac{1}{\rho} \mathbbm{1}_n + \frac{1}{n} \sum_{\mu = 1}^m \frac{\partial_\omega g_\mathrm{out}(y_\mu,0,\rho \langle \lambda \rangle_\nu/\alpha)}{1 + \frac{\rho \langle \lambda \rangle_\nu}{\alpha}\partial_\omega g_\mathrm{out}(y_\mu,0,\rho \langle \lambda \rangle_\nu/\alpha)} \overline{\Phi_{\mu i}} \Phi_{\mu j}.
\end{align*}

\subsection{Unification of the approaches}\label{subsec:unification}

We now detail our main claims and results concerning the spectral methods we just derived.

\medskip
\noindent
\textbf{The optimal spectral method and the Bethe Hessian}

\medskip
\noindent
Very importantly, as opposed to previous approaches, our derivation is \emph{constructive}: we start from the fully-explicit expression of the TAP free entropy given 
in eq.~\eqref{eq:TAP_free_entropy} and simply compute its Hessian at the trivial fixed point. 
From the statistical physics literature (as we detailed in Section~\ref{subsec:main_results}), we expect that the optimal spectral method will be given by the largest eigenvalue (and associated eigenvector)
of this Hessian.
The result of our computation of this Hessian was given in eq.~\eqref{eq:def_MTAP}.
Importantly, this implies that the optimal spectral method that can be built from the data $\bPhi$ and the observations $\{y_\mu\}$ belongs to the class of methods given by eq.~\eqref{eq:general_class_spectral_methods}.
Our conjecture therefore also gives weight to many previous analysis of spectral methods for phase retrieval, which only studied spectral methods of the type of eq.~\eqref{eq:general_class_spectral_methods} \cite{lu2020phase,mondelli2019fundamental,luo2019optimal,ma2019spectral}.

\medskip
\noindent
\textbf{Relating linearized-AMP and the Bethe Hessian}

\medskip
\noindent
Our derivation of $\bM^{\mathrm{(LAMP)}}$ is \emph{constructive} as well, and in this sense fundamentally differs from the L-AMP algorithms designed in \cite{ma2019spectral} 
to assess the performance of other spectral methods. 
We start by a proposition, proven in Appendix~\ref{subsec_app:equivalence_lamp_tap}, which relates the eigenpairs of the two methods.
\begin{proposition}\label{prop:relating_lamp_tap}
  Without loss of generality, we assume $\langle \lambda \rangle_\nu = \alpha$. 
 Let $z_\mu \equiv \partial_\omega g_\mathrm{out}(y_\mu,0,\rho \langle \lambda \rangle_\nu/\alpha)$, and $(\lambda_\mathrm{LAMP},\bv)$ be an eigenpair of $\bM^\mathrm{(LAMP)}$. Assume
 that $\lambda_\mathrm{LAMP} + \rho z_\mu \neq 0$ for all $\mu = 1,\cdots,m$. Then $\bPhi^\dagger \mathrm{Diag}(z_\mu) \bv \neq 0$, and we let $\bxhat \propto \bPhi^\dagger \mathrm{Diag}(z_\mu) \bv$ with $\lVert \bxhat \rVert^2 = n$. Moreover:
 \begin{align*}
    \Big\{\frac{1}{m} \sum_{\mu=1}^m \frac{\rho z_\mu}{\lambda_\mathrm{LAMP} + \rho z_\mu} \bPhi_\mu \bPhi_\mu^\dagger \Big\} \bxhat &= \bxhat.
 \end{align*}
 Conversely, let $\bx$ be an eigenvector of $\bM^\mathrm{(TAP)}$ with norm $\lVert\bx\rVert^2 = n$, with associated eigenvalue $\lambda_\mathrm{TAP}$.
 We define $\bu \equiv \mathrm{Diag}[(1+\rho z_\mu)^{-1}] \bPhi \bx / \sqrt{n}$. Then one has:
 \begin{align*}
 \bM^\mathrm{(LAMP)} \bu &=  \bu + \rho \lambda_\mathrm{TAP} \mathrm{Diag}(1+\rho \partial_\omega g_\mathrm{out}(y_\mu, 0, \rho)) \bu.
 \end{align*}
 Moreover, if $\lambda_\mathrm{TAP} = 0$, eq.~\eqref{eq:correspondance_LAMP_TAP_estimators} applied to $\bu$ yields the same performance as the TAP estimator.
\end{proposition}
\noindent
By considering $\lambda_\mathrm{LAMP} = 1$ and $\lambda_\mathrm{TAP} = 0$, one immediately deduces two important consequences of Proposition~\ref{prop:relating_lamp_tap} and the definitions of the methods (cf.\ eqs.~\eqref{eq:M_LAMP_gout},\eqref{eq:def_MTAP}):
\begin{itemize}[leftmargin=*]
  \item The appearance of an unstable direction, in the spectrum of $\bM^\mathrm{(TAP)}$ (i.e. a positive eigenvalue) and of $\bM^\mathrm{(LAMP)}$ (i.e. an eigenvalue with real part greater than $1$),
  occurs at a common threshold (i.e.\ the \emph{weak-recovery} threshold, given by eq.~\eqref{eq:wr_threshold_general}).
  \item An eigenvalue $0$ appears in the spectrum of $\bM^\mathrm{(TAP)}$ if and only if an eigenvalue $1$ appears in the spectrum of 
  $\bM^\mathrm{(LAMP)}$. These two eigenvalues therefore correspond to \emph{marginal stability} of the linear dynamics. Moreover, the two estimators 
  associated to these eigenvalues are identical, i.e.\ $\bM^\mathrm{(LAMP)}$ \emph{contains the optimal estimator}.
  Importantly, this estimator is different from the largest eigenvector of $\bM^\mathrm{(LAMP)}$, which reaches only suboptimal performance 
  as we will see in Section~\ref{sec:numerics}.
\end{itemize}
\section{Numerical experiments and perspectives}\label{sec:numerics}

In this section, we numerically assess our predictions and compare the performance of the spectral methods on various problems.
In Section~\ref{subsec:performance_methods}, we consider the recovery of a randomly generated signal 
with different right-rotationally invariant sensing matrix ensembles.
In Sec.~\ref{subsec:transition_spectra}, we illustrate the transition phenomena in the spectra of $\bM^\mathrm{(TAP)}$ and $\bM^\mathrm{(LAMP)}$, which 
raise interesting random matrix theory questions.
Finally, in Section~\ref{subsec:real_image}, we validate our predictions for the recovery of a natural image with various matrix ensembles. We numerically verify that 
all our conclusions derived for random signals still hold in this setting. 
The numerical code used to generate all figures is available in the supplementary material.

\noindent
\textbf{Another spectral method --} In the figures, we sometimes consider another spectral method, called $\bM^\mathrm{(MM)}$. It is obtained by naively considering the preprocessing 
function of \cite{mondelli2019fundamental}, which was shown to achieve the optimal transition for Gaussian sensing matrices. 
More precisely, we have (assuming $\rho = 1$ and $\langle \lambda \rangle_\nu = \alpha$):
$\mathcal{T}_{\mathrm{MM}}(y) \equiv \partial_\omega g_\mathrm{out}(y,0,1) / \Big[\sqrt{\frac{2 \alpha}{\beta}} + \partial_\omega g_\mathrm{out}(y,0,1) \Big]$.
In particular note that at $\alpha = \beta/2$, we have $\mathcal{T}_\mathrm{MM} = \mathcal{T}^\star$, so that $\mathcal{T}_\mathrm{MM}$ indeed achieves the optimal weak-recovery transition for Gaussian matrices, 
for which $\alpha_\mathrm{WR,Algo} = \beta / 2$.

\subsection{Performance of the spectral methods}\label{subsec:performance_methods}

\begin{figure}[ht]
   \centering
    \begin{subfigure}[b]{0.95\textwidth}
        \centering
       \includegraphics[width=\textwidth]{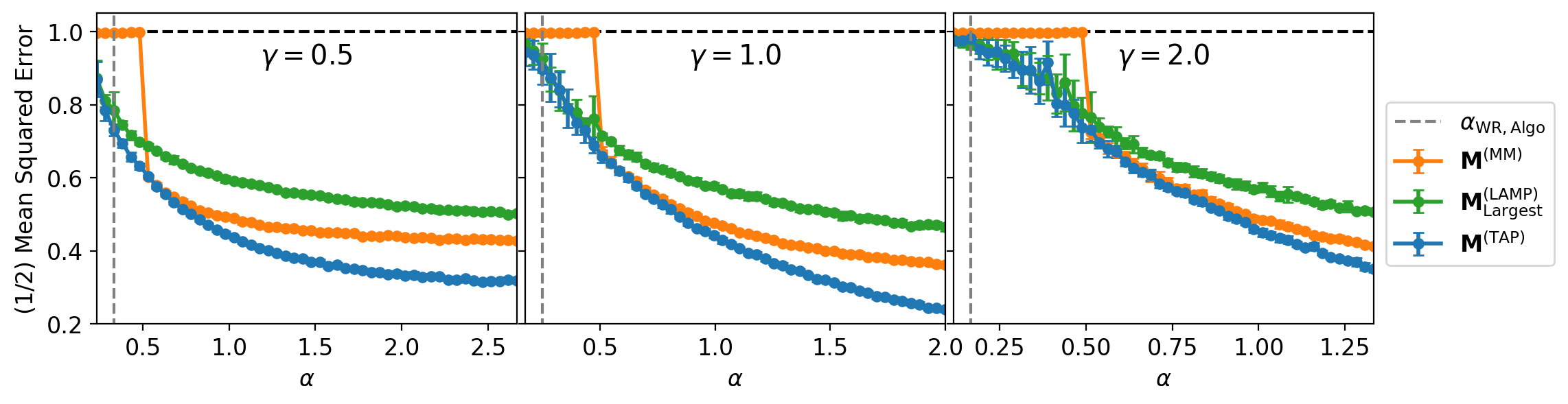}
       \caption{
        Product of two real i.i.d.\ Gaussian sensing matrices with a size ratio $\gamma \in \{0.5,1.0,2.0\}$.
        The simulations were done using $m = 10000$, and error bars are taken over $10$ instances.
      \label{fig:mse_product_real_gaussian}
       }
    \end{subfigure}
    \begin{subfigure}[b]{0.95\textwidth}
      \centering
       \includegraphics[width=0.95\textwidth]{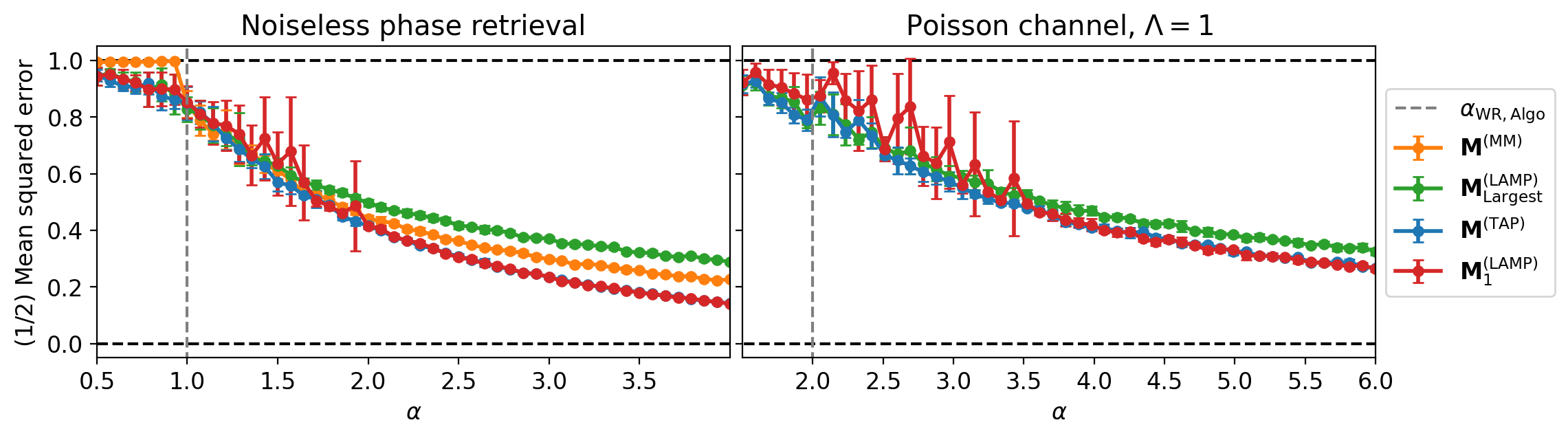}
      \caption{Complex Gaussian matrix, in noiseless phase retrieval and in Poisson-noise phase retrieval with $\Lambda = 1$.
    The simulations were done using $m = 10000$ (noiseless case), $12000$ (Poisson case), and the error bars are taken over $10$ (noiseless case), $5$ (Poisson case)
    instances.\label{fig:mse_complex_gaussian_noiseless_poisson}
       }
    \end{subfigure}
  \caption{
    Mean squared error achieved by the different spectral methods in two different settings.
    \label{fig:mse_plots}
  }
\end{figure}

We show the performance of the spectral methods to recover a random signal in three different cases, that we briefly describe:
\begin{itemize}[leftmargin = *]
    \item In Fig.~\ref{fig:mse_orthogonal}, we consider noiseless real phase retrieval (i.e.\ sign retrieval), 
    with uniformly sampled column-unitary sensing matrices. We also show that our conclusions transfer to randomly subsampled Hadamard 
    matrices, validating the conclusions of \cite{dudeja2020universality}.
    \begin{figure}
      \centering
    \includegraphics[width=0.8\textwidth]{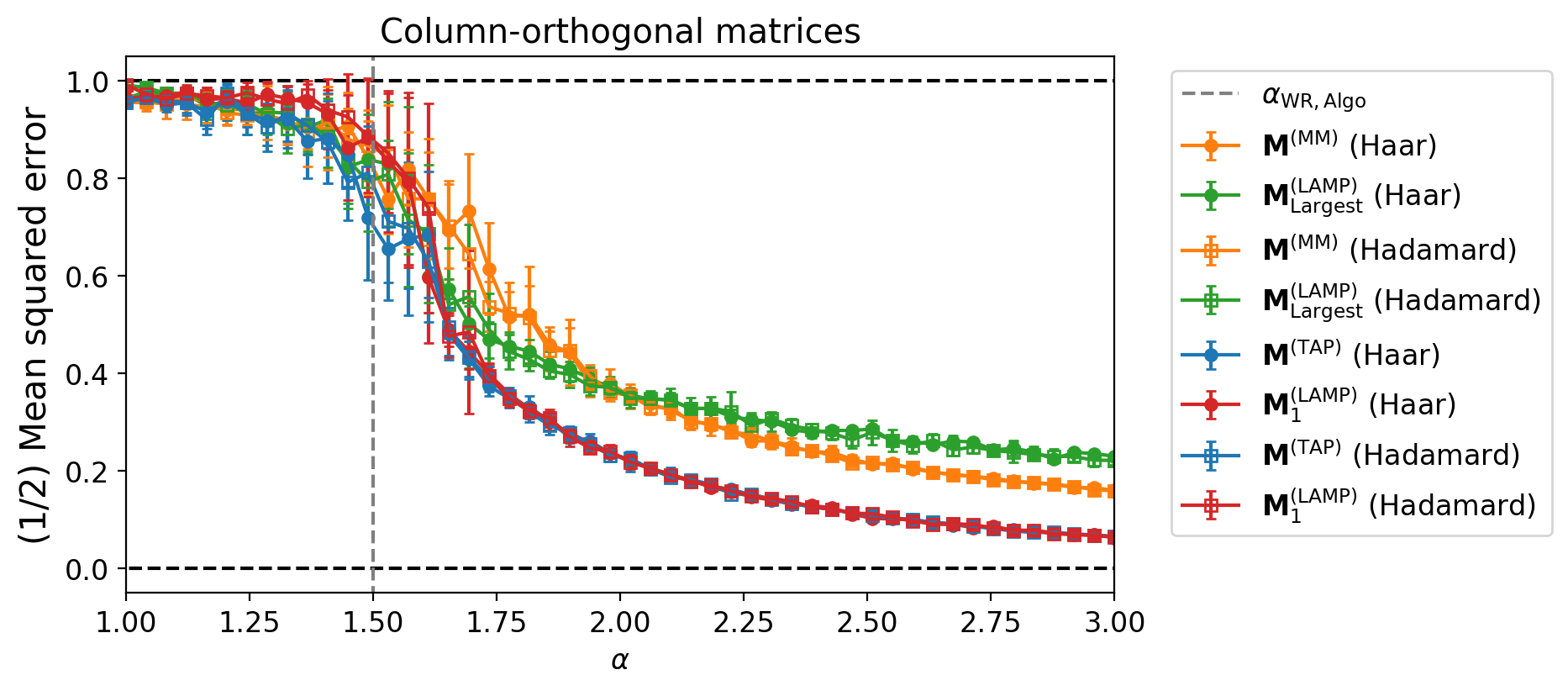}
    \caption{
    Mean squared error achieved by our spectral methods and a naive version of the spectral method of \cite{mondelli2019fundamental} for real column-orthogonal sensing matrices
    and a noiseless channel. We give the performance on uniformly sampled column-orthogonal matrices as well as randomly subsampled Hadamard matrices. The 
    simulations were done using $m = 8192$, and the error bars are taken over $10$ instances.
    }
    \label{fig:mse_orthogonal}
    \end{figure}
    \item In Fig.~\ref{fig:mse_product_real_gaussian} we consider noiseless real phase retrieval when the sensing matrix is a product of two Gaussian i.i.d.\ 
    matrices. This setup can for instance be interpreted as Gaussian phase retrieval in which the signal is drawn from a known generative prior, 
    similarly to the analysis of \cite{aubin2020exact}. Importantly, it is not covered by any previous analysis of the spectral methods, 
    emphasizing the generality of the framework of Hypothesis~\ref{hyp:sensing_matrix}.
    \item In Fig.~\ref{fig:mse_complex_gaussian_noiseless_poisson}, we compare our results in noiseless and noisy settings. 
    More precisely, we consider complex phase retrieval with a Gaussian sensing matrix, and either a noiseless channel or a Poisson observation channel with intensity $\Lambda > 0$:
    \begin{align*}
      P_\mathrm{out}(y|z) = e^{-\Lambda |z|^2}\sum_{k=0}^\infty \delta(y - k) \frac{\Lambda^k |z|^{2k}}{k!}.
    \end{align*}
    This latter channel is particularly relevant for optical applications, in which the detector can be modeled as being affected by a Poisson noise.
    In both cases, we find that all our conclusions on the optimality of the $\bM^\mathrm{(TAP)}$, and on the link between 
    $\bM^\mathrm{(LAMP)}$ and $\bM^\mathrm{(TAP)}$, still hold.
\end{itemize}

\subsection{Transition phenomena in the spectra}\label{subsec:transition_spectra}

We illustrate the weak-recovery transition in the spectra of the different methods.
\begin{figure}[t]
  \centering
\includegraphics[width=0.6\textwidth]{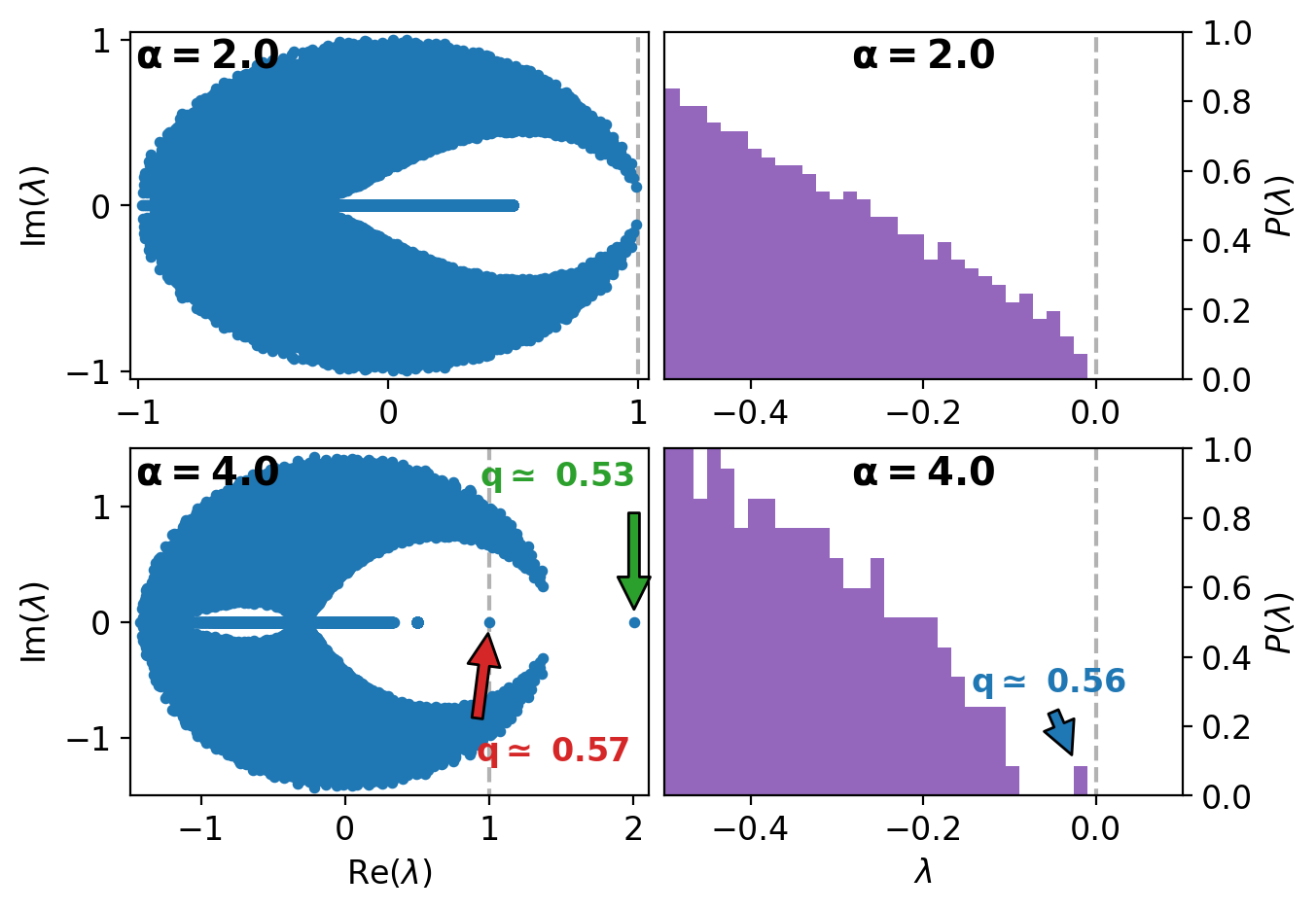}
\caption{Transition in the spectra of $\bM^{\mathrm{(LAMP)}}$ (left) and $\bM^{\mathrm{(TAP)}}$ (right) for a complex Gaussian $\bPhi$ and a Poisson channel with $\Lambda = 1$. For $\alpha > \alpha_\mathrm{WR,Algo} = 2$, we indicate the approximate overlap $q$ corresponding to the the relevant eigenvalues.
}
\label{fig:transition_complex_gaussian_poisson}
\end{figure}
Precisely, we confirm the following claims of Section~\ref{subsec:unification}:
\begin{itemize}[leftmargin=*]
  \item Both $\bM^\mathrm{(LAMP)}$ and $\bM^\mathrm{(TAP)}$ have a largest eigenvalue (in real part) that detaches from the bulk for $\alpha > \alpha_\mathrm{WR,Algo}$, given by eq.~\eqref{eq:wr_threshold_general}.
  \item In the regime in which weak-recovery is possible, the largest eigenvalue of $\bM^\mathrm{(TAP)}$ approaches $0$ as $n \to \infty$. The associated eigenvector achieves optimal correlation with the signal (among spectral methods)
  as $n \to \infty$.
  \item $\bM^\mathrm{(LAMP)}$ gives \emph{two} estimators that are positively correlated with the signal for $\alpha > \alpha_\mathrm{WR,Algo}$. The first one corresponds to its largest eigenvalue in real part, 
  and achieves worse correlation than the largest eigenvector of $\bM^\mathrm{(TAP)}$. The second one corresponds to an eigenvalue inside the bulk (but isolated from the other eigenvalues) that approaches $1$ as $n \to \infty$, and achieves the same optimal performance as the estimator given by $\bM^\mathrm{(TAP)}$.
\end{itemize}
We verify these claims for different values of $\alpha$, below and above the weak-recovery 
threshold, in complex Gaussian phase retrieval with Poisson-noise, in Fig.~\ref{fig:transition_complex_gaussian_poisson}.
We complete this analysis in Appendix~\ref{sec:app_numerics}, by considering noiseless phase retrieval and more values of $\alpha$ in Fig.~\ref{fig:transition_complex_gaussian_noiseless}, 
and product of complex Gaussian matrices and structured signals in Fig.~\ref{fig:transition_real_image}.

\noindent
\textbf{Remark --} In the shown figures there is a very small discrepancy between the overlaps achieved by the principal eigenvector of $\bM^{\mathrm{(TAP)}}$
and the eigenvector of $\bM^{\mathrm{(LAMP)}}$ with eigenvalue $1$. This is due to the fact that the subplots of Fig.~\ref{fig:transition_complex_gaussian_poisson}
(and similarly for Fig.~\ref{fig:transition_complex_gaussian_noiseless})
are generated with different instances of the matrix $\bPhi$ and signal $\bX^\star$.

\noindent
\textbf{On the performance of the spectral methods --} When weak recovery is possible the largest eigenvalue of $\bM^{\mathrm{(TAP)}}$ concentrates on $0$ as we 
noticed. However, the spectrum of $\bM^{\mathrm{(TAP)}}$ also contains many very large negative eigenvalues.
In practice, we use an inverse iteration method to quickly estimate the associated eigenvector.
We use a similar approach for $\bM^{\mathrm{(LAMP)}}$, using inverse iterations to estimate the eigenvector with eigenvalue $1$, and 
usual power iterations for the largest eigenvalue.

\subsection{Real image reconstruction}\label{subsec:real_image}

\begin{figure}
  \centering
\includegraphics[width=\textwidth]{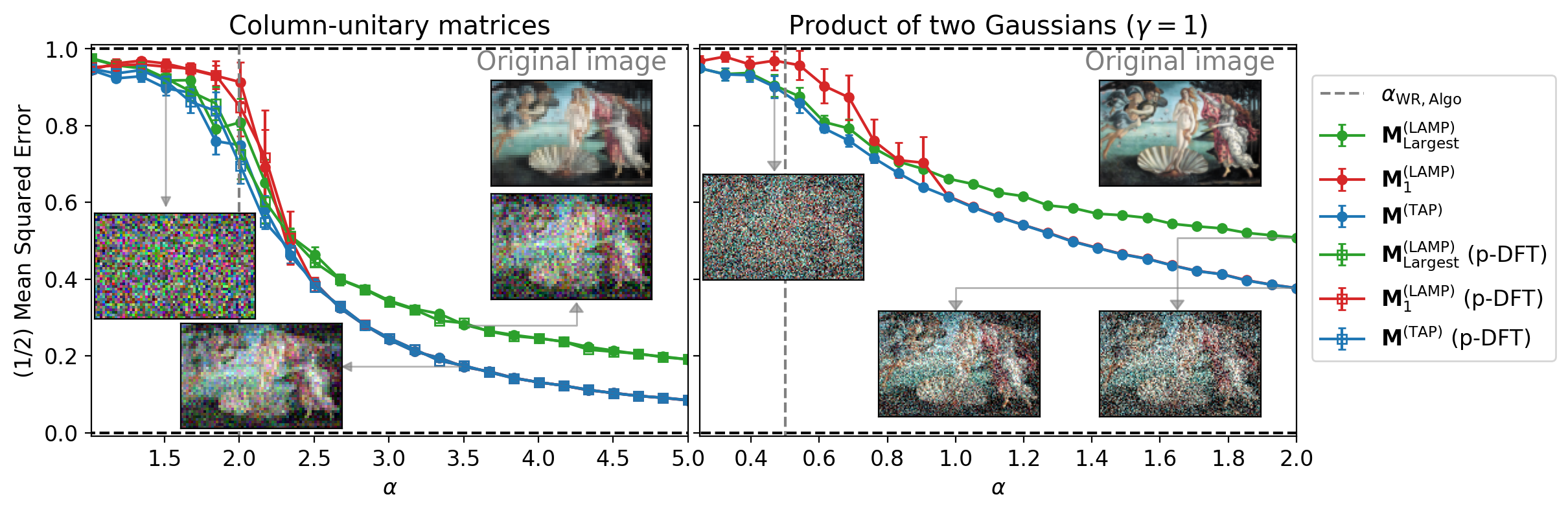}
\caption{Mean squared error achieved by the different spectral methods for the recovery of a natural image in noiseless phase retrieval. 
We consider column-unitary matrices $\bPhi$ (both uniformly sampled and partial DFT matrices, left) and 
the product of two complex Gaussian matrices with aspect ratio $\gamma = 1$ (right). 
We reduced each dimension of the original $1280 \times 820$ image by a factor $20$ (left) or $10$ (right), and we average the MSE over $5$ instances and the $3$ RGB channels (which are recovered independently). 
}
\label{fig:mse_real_image}
\end{figure}
As a final analysis, we numerically investigate our predictions for the reconstruction of a natural image. For comparability, we 
consider the image of \emph{The Birth of Venus} already used in \cite{mondelli2019fundamental,ma2019spectral}.
Although this signal is not i.i.d., we will see that all our previous conclusions, numerically investigated in Sections~\ref{subsec:performance_methods},\ref{subsec:transition_spectra},
transfer to this case.
We consider a noiseless phase retrieval channel and different sensing matrices $\bPhi$: multiple ensembles of column-unitary matrices (which partly reproduces the analysis of \cite{ma2019spectral}) and a product of two complex Gaussian matrices with aspect ratio $\gamma = 1$.
In particular we consider partial DFT matrices, introduced in \cite{ma2014turbo,ma2019spectral}, which are an ensemble of column-unitary matrices obtained from the usual DFT matrices.
Namely, there are defined for $m \geq n$ as $\bPhi/\sqrt{n} = \bF \bS \bP$, with $\bF \in \bbC^{m \times m}$ a DFT matrix, $\bS \in \bbR^{m \times n}$
containing $n$ columns (randomly taken) of the identity matrix $\mathds{1}_m$, and $\bP$ a diagonal of random phases.
In Fig.~\ref{fig:mse_real_image}, we give the MSE obtained by the different spectral methods and these two matrix ensembles. 
We also give examples of the images recovered by the algorithms.
Eventually, despite the fact that the signal (and possibly the matrix as well) is structured,
we still observe the same transition phenomena in the spectra of $\bM^\mathrm{(TAP)}$ and $\bM^\mathrm{(LAMP)}$, as 
shown in the supplementary material, in Fig.~\ref{fig:transition_real_image}. Namely, we still observe that the optimal estimator is associated with marginal stability of both spectral methods,
while the largest eigenvalue of $\bM^\mathrm{(LAMP)}$ is associated to a non-optimal estimator.

\begin{figure}[ht]
  \centering
  \includegraphics[width=0.8\textwidth]{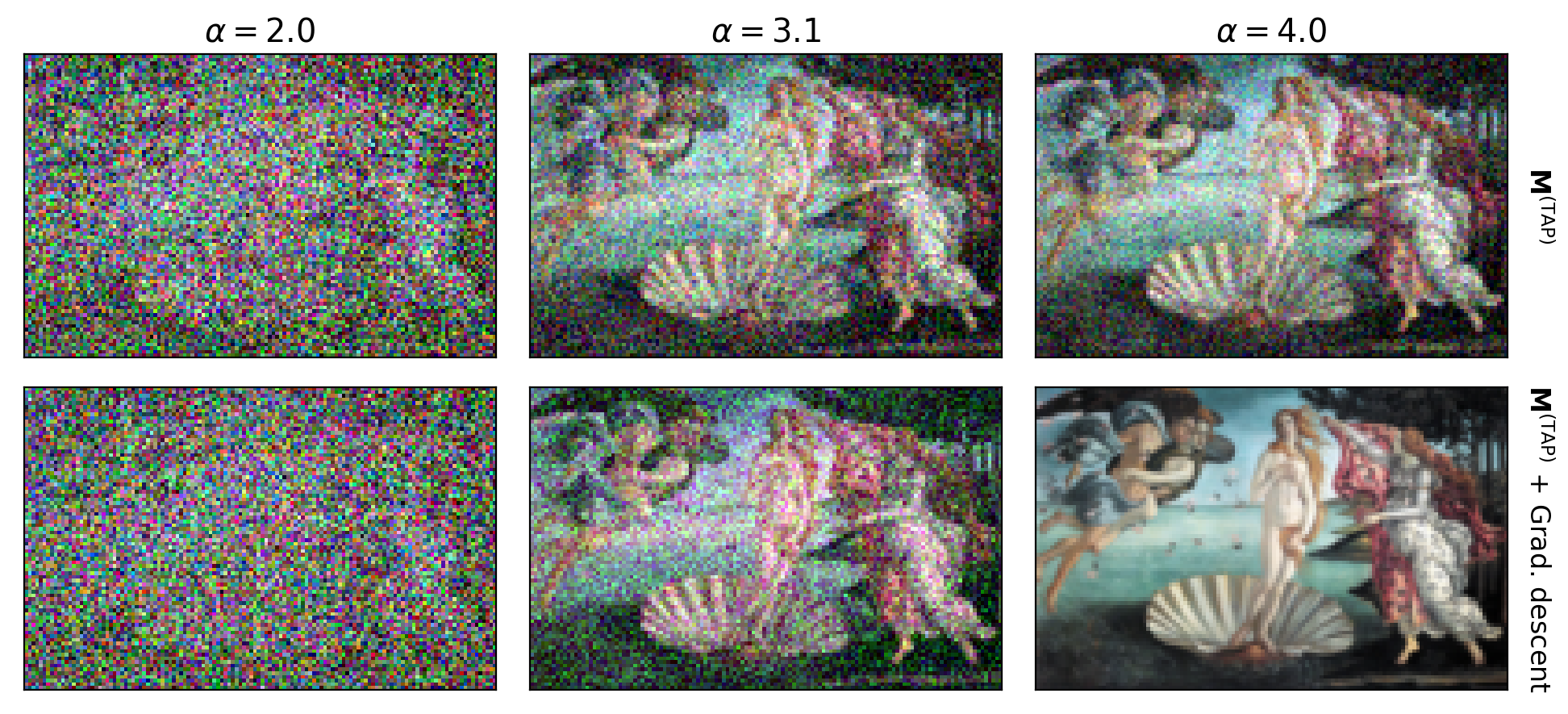}
  \caption{
    Reconstruction of a real image in noiseless phase retrieval with partial DFT matrices. 
    We reduce the image size from $1280 \times 820$ to $128 \times 82$.
    We compare, for three different values of $\alpha$, the 
    estimators of $\bM(\mathcal{T}^\star)$ (top line) and the estimator obtained by running a gradient descent procedure starting from the estimator of $\bM(\mathcal{T}^\star)$ (bottom line).
    We recover the 3 RGB channels with independent instances of the sensing matrix.
  }
  \label{fig:comparison_TAP_with_GD}
\end{figure}

Let us also illustrate how this spectral method can be combined with a subsequent local optimization algorithm.
We use the spectral estimator as the initialization point to running vanilla gradient descent on the square loss $L(\bx) \equiv \frac{1}{2m} \sum_{\mu=1}^m \Big\{\big|\frac{(\bPhi \bx)_\mu}{\sqrt{n}}\big|^2 - \big|\frac{(\bPhi \bX^\star)_\mu}{\sqrt{n}}\big|^2\Big\}^2$.
This allows to already obtain a perfect recovery of the image for $\alpha = 4$, as shown in Fig.~\ref{fig:comparison_TAP_with_GD}.
In Appendix~\ref{subsec:app_performance_spectral_GD} we expand this analysis by showing the MSE achieved by the gradient descent procedure. In particular, we confirm that combining the gradient descent with the spectral initialization 
allows to reach perfect recovery at finite $\alpha$, which is not possible with the ``vanilla'' spectral methods.

\subsection{Perspectives}
Our analysis raises interesting open questions, both from the random matrix theory and the statistical physics viewpoint.
\begin{itemize}[leftmargin=*]
  \item First, we notice that the optimal estimator is always associated with \emph{marginal stability}, both in $\bM^{\mathrm{(LAMP)}}$ and $\bM^{\mathrm{(TAP)}}$.
A clear understanding of this marginal stability is still lacking. 
Note that this marginal stability was already observed in \cite{ma2019spectral} for phase retrieval with column-unitary matrices, 
and in the context of community detection, a marginally-stable eigenvalue inside the bulk of the non-backtracking operator was already observed to be associated with Bayes-optimality 
in \cite{dall2019revisiting}\footnote{This was even used as a criteria to evaluate the Nishimori temperature from the Bethe Hessian in \cite{dall2021nishimori}, which appeared a few months after our analysis.}.
Moreover, the principal eigenvector of the matrix $\bM^{\mathrm{(LAMP)}}$ is associated to an \emph{unstable} direction, thus dominating the dynamics of the linearized-AMP. 
However its achieved correlation is smaller than the one achieved by the marginally stable, optimal, eigenvector. 
We also noticed that the eigenvectors of $\bM^{\mathrm{(TAP)}}$ \emph{do not contain any information about this suboptimal estimator}\footnote{
In particular, this is an important distinction between our L-AMP constructive derivation and the L-AMP algorithms of \cite{ma2019spectral}, which are \emph{designed} to match the 
spectral methods of the type $\bM(\mathcal{T})$: in the latter, it was shown that the L-AMP estimator always matched the one of the spectral method.
}.
This blindness of $\bM^{\mathrm{(TAP)}}$ to the principal eigenvector of $\bM^{\mathrm{(LAMP)}}$ is very puzzling from a theoretical point of view.
Indeed, as shown in \cite{maillard2019high} and reminded in Section~\ref{subsubsec:tap_free_entropy}, the stationary limit of G-VAMP (Algorithm~\ref{algo:gvamp}) is in exact correspondence with 
the stationary point equations of the TAP free entropy. One would therefore expect the two spectral methods $\bM^\mathrm{(LAMP)}$ and $\bM^\mathrm{(TAP)}$ to contain the same physical information on the system.
Finally, the different qualitative behaviors of the two methods (instability of $\bM^{\mathrm{(LAMP)}}$ a opposed to marginal stability of $\bM^{\mathrm{(TAP)}}$) only deepens this puzzle, and understanding this disparity is an interesting open problem.
\item Importantly, our analysis is essentially not rigorous (hence the use of conjectures). 
An interesting perspective would be to establish rigorously our statements,
in similarity with what is proven in \cite{dudeja2020analysis} on the analysis of \cite{ma2019spectral} for column-unitary matrices. 
This would require a random matrix theory analysis of the ``BBP''\footnote{i.e.\ the appearance of a largest eigenvalue detached from the bulk of the other eigenvalues, as $\alpha$ increases.
It was first rigorously analyzed in \cite{baik2005phase} for spiked Gaussian matrices.} transition in matrices of the form of eq.~\eqref{eq:def_MTAP}, which is, to 
the best of our knowledge, lacking in the generic rotationally-invariant case. Another approach would be to use the (rigorously known)
\emph{state evolution} (SE) of AMP, which allows to track its asymptotic performance.
This approach was considered in \cite{ma2019spectral,dudeja2020universality}: importantly, this method also provides analytically the asymptotic performance of the spectral method, 
which is not derived in the present work.
\item 
Another important perspective is to apply our methods in real-world settings in which the way the data and the signal are generated is not necessarily known.
Our analysis of a real image (cf Fig.~\ref{fig:mse_real_image}) suggests that having a structured prior distribution does not harm our conclusions. 
The influence of a so-called ``mismatched'' setting in the channel distribution (i.e.\ when the data is generated with a distribution $P_\mathrm{out}^0$ and inferred 
with a different distribution $P_\mathrm{out}$) is however less clear, and we leave it for future work.
\end{itemize}

\section*{Acknowledgements}

Funding is acknowledged by AM from ``Chaire de
recherche sur les mod\`eles et sciences des donn\'ees'', Fondation CFM pour la Recherche-ENS.  This work is supported by the ERC under the European Union’s Horizon 2020 Research and Innovation Program
714608-SMiLe, as well as by the French Agence Nationale de la Recherche under grant ANR-17-CE23-0023-01
PAIL and ANR-19-P3IA-0001 PRAIRIE.
Part of this work was done when Yue M. Lu
was visiting Ecole Normale as a CFM-ENS “Laplace” invited researcher.

\bibliographystyle{alpha}
\bibliography{refs}

\newpage
\appendix

\section{Linearized Approximate Message Passing in the complex case}\label{sec_app:lamp_derivation}

In the complex case, we write the linearization of Algorithm~\ref{algo:gvamp} as:
\begin{align}\label{eq:linearization_trivial_fixed_point_vector_complex}
  \begin{cases}
    \delta \hat{\bx}_1^t = \nabla_{\bT} g_{x1}(0,0) \delta \bT_1^t + \nabla_{\bar{\bT}} g_{x1}(0,0) \overline{\delta \bT_1^t}, & \\
    \delta \hat{\bz}_1^t = \nabla_{\bR} g_{z1}(0,\rho^{-1}) \delta \bR_1^t + \nabla_{\bar{\bR}} g_{z1}(0,\rho^{-1}) \overline{\delta \bR_1^t}, & \\
    \delta \bT_2^t = \frac{1}{\rho} \delta \hat{\bx}_1^t - \delta \bT_1^t, & \delta \bR_2^t = \frac{\alpha}{\rho \langle \lambda \rangle_\nu} \delta \hat{\bz}_1^t - \delta \bR_1^t, \\
    \delta \hat{\bx}_2^t = \nabla_{\bT} g_{x2}(0,0,\rho^{-1},0) \delta \bT_2^t  + \nabla_{\bR} g_{x2}(0,0,\rho^{-1},0) \delta \bR_2^t, & \\ 
    \delta \hat{\bz}_2^t = \nabla_{\bT} g_{z2}(0,0,\rho^{-1},0) \delta \bT_2^t  + \nabla_{\bR} g_{z2}(0,0,\rho^{-1},0) \delta \bR_2^t, & \\
    \delta \bT_1^{t+1} = \frac{1}{\rho} \delta \hat{\bx}_2^t - \delta \bT_2^t, & \delta \bR_1^{t+1} = \frac{\alpha}{\rho \langle \lambda \rangle_\nu} \delta \hat{\bz}_2^t - \delta \bR_2^t.
  \end{cases}
\end{align}
Recall that here $\partial_z,\partial_{\bar{z}}$ are the usual Wirtinger derivatives.
Since the functions $g_{x2},g_{z2}$, defined in eq.~\eqref{eq:auxiliary_functions_gvamp}, are obviously holomorphic, we did not include their derivative $\partial_{\bar{z}}$ as it is 
trivially zero.
Moreover, we assumed that $P_0(z),P_\mathrm{out}(y|z)$ are functions of $|z|^2$ (i.e.\ \emph{spherical symmetry}), which defined our phase retrieval problem.
Starting from the definition of eq.~\eqref{eq:auxiliary_functions_gvamp}, this implies that 
\begin{align*}
    \partial_{\overline{T_i}} g_{x1}(0,0) = 2 \big(\EE_{P_0}[z^2] - \EE_{P_0}[z]\big) = 0,
\end{align*}
in which the last equality is a consequence of the spherical symmetry. In the same way, one obtains $\nabla_{\bar{\bR}} g_{z1}(0,\rho^{-1}) = 0$.
We can then compute, as in the real case (cf eq.~\eqref{eq:auxiliary_functions_trivial_fixed_point}):
\begin{align}
  \begin{cases}\label{eq:auxiliary_functions_trivial_fixed_point_complex}
  \partial_{T_j} [(g_{x1}(0,0)_i] &= \rho \, \delta_{ij}, \\
  \partial_{R_\nu} [g_{z1}(0,\rho^{-1})_\mu] &= \delta_{\mu \nu} \EE_{P_\mathrm{out}(y_\mu,0,\rho\langle \lambda \rangle_\nu / \alpha)} \equiv \delta_{\mu \nu} v(y_\mu), \\
  \partial_{T_j} [g_{x2}(0,0,\rho^{-1},0)] &= \rho \, \delta_{ij}, \\
  \partial_{R_\mu} [g_{x2}(0,0,\rho^{-1},0)_i] &= \rho (\bV \bS^\intercal \bU^\dagger)_{i \mu} = \rho \frac{(\bPhi^\dagger)_{i \mu}}{\sqrt{n}}, \\
  \partial_{T_i} [g_{z2}(0,0,\rho^{-1},0)_\mu] &= \rho \, \frac{\Phi_{\mu i}}{\sqrt{n}}, \\
  \partial_{R_\nu} [g_{z2}(0,0,\rho^{-1},0)_\mu] &= \rho \, \frac{(\bPhi \bPhi^\dagger)_{\mu \nu}}{n}.
  \end{cases}
\end{align}
The derivation of the real case then straightforwardly transfers to the complex case, and we reach eq.~\eqref{eq:linearization_gvamp_final}
in the complex case, as claimed.
\section{The Hessian of the TAP free entropy}\label{sec:app_hessian_tap}

\subsection{The derivatives of the parameters at the trivial fixed point}\label{subsec:app_derivatives_trivial_point_tap}

We start from the relations of eq.~\eqref{eq:tap_eqs_complete}. 
Let us differentiate them with respect to $m_i^{(a)}$, for any $a \in \{1,\beta\}$ and $i \in \{1,\cdots,n\}$.
We denote $P_\mathrm{out}^\mu \equiv P_\mathrm{out}(y_\mu,\omega_\mu,b)$.
We get after tedious calculations the cumbersome equations (valid for any $\bmm$) :
\begin{subnumcases}{\label{eq:derivative_tap_eqs_mi}}
     \delta_{ij} \bbe_a =\beta\Big\{-\frac{1}{2}\frac{\partial\gamma}{\partial m_i^{(a)}}(\EE_{P_0(\gamma,\lambda_j)}[x|x|^2] - \EE_{P_0(\gamma,\lambda_j)}[x]\EE_{P_0(\gamma,\lambda_j)}[|x|^2]) \\ 
     \nonumber \hspace{1cm}- \EE_{P_0(\gamma,\lambda_j)}\Big[x \big(x \cdot \frac{\partial \lambda_j}{\partial m_i^{(a)}}\big) \Big]  + \EE_{P_0(\gamma,\lambda_j)}[x]  \EE_{P_0(\gamma,\lambda_j)}\big[x \cdot \frac{\partial \lambda_j}{\partial m_i^{(a)}}\big] \Big\}, & \\
     \frac{\partial \sigma^2}{\partial m_i^{(a)}} = \frac{1}{n} \sum_{j=1}^n \Big[-2 m_j^{(a)} \delta_{ij} + \beta \Big\{-\frac{1}{2}\frac{\partial\gamma}{\partial m_i^{(a)}}(\EE_{P_0(\gamma,\lambda_j)}[|x|^4] - (\EE_{P_0(\gamma,\lambda_j)}[|x|^2])^2) \\ 
     \nonumber \hspace{1cm}+ \EE_{P_0(\gamma,\lambda_j)}\Big[|x|^2 \big(x \cdot \frac{\partial \lambda_j}{\partial m_i^{(a)}}\big) \Big]  - \EE_{P_0(\gamma,\lambda_j)}[|x|^2]  \EE_{P_0(\gamma,\lambda_j)}\big[x \cdot \frac{\partial \lambda_j}{\partial m_i^{(a)}}\big]
     \Big\}\Big], &\\ 
     \frac{\partial g_\mu}{\partial m_i^{(a)}} = \frac{1}{b}\frac{\partial \omega_\mu}{\partial m_i^{(a)}} + \frac{\partial b}{\partial m_i^{(a)}} \Big\{b^{-2} \EE_{P_\mathrm{out}^\mu}[h-\omega_\mu] \\ 
     \nonumber + \frac{\beta}{2 b^3} \big(\EE_{P_\mathrm{out}^\mu}[(h-\omega_\mu) |h-\omega_\mu|^2]- \EE_{P_\mathrm{out}^\mu}[h-\omega_\mu] \EE_{P_\mathrm{out}^\mu}[|h-\omega_\mu|^2]\big) \Big\} & \\
     \nonumber  - \frac{\beta}{b^2} \Big(\EE_{P_\mathrm{out}^\mu}\big[(h-\omega_\mu) (h-\omega_\mu) \cdot \frac{\partial \omega_\mu}{\partial m_i^{(a)}}\big]- \EE_{P_\mathrm{out}^\mu}[h-\omega_\mu]\EE_{P_\mathrm{out}^\mu}\big[(h-\omega_\mu) \cdot \frac{\partial \omega_\mu}{\partial m_i^{(a)}}\big]\Big) ,& \\
     \frac{\partial r}{\partial m_i^{(a)}} = \frac{1}{m} \sum_{\mu=1}^m \Big\{2 g_\mu \cdot \frac{\partial g_\mu}{\partial m_i^{(a)}} - b^{-2} \frac{\partial b}{\partial m_i^{(a)}}  
     + \frac{1}{b^2}\EE_{P_\mathrm{out}^\mu}\big[(h-\omega_\mu) \cdot \frac{\partial \omega_\mu}{\partial m_i^{(a)}}\big] & \\ 
     \nonumber + \frac{\partial b}{\partial m_i^{(a)}} \Big\{2 b^{-3} \EE_{P_\mathrm{out}^\mu}[|h-\omega_\mu|^2] + \frac{\beta}{2 b^4} \big(\EE_{P_\mathrm{out}^\mu}[|h-\omega_\mu|^4]- \big(\EE_{P_\mathrm{out}^\mu}[|h-\omega_\mu|^2]\big)^2\big) \Big\} & \\
     \nonumber  + \frac{\beta}{b^3} \Big(\EE_{P_\mathrm{out}^\mu}\big[|h-\omega_\mu|^2 (h-\omega_\mu) \cdot \frac{\partial \omega_\mu}{\partial m_i^{(a)}}\big]- \EE_{P_\mathrm{out}^\mu}[|h-\omega_\mu|^2]\EE_{P_\mathrm{out}^\mu}\big[(h-\omega_\mu) \cdot \frac{\partial \omega_\mu}{\partial m_i^{(a)}}\big]\Big) \Big\}, & \\ 
     \label{eq:derivative_tap_gamma}
     \frac{\partial \gamma}{\partial m_i^{(a)}} = - 2\Big[ \frac{\partial \sigma^2}{\partial m_i^{(a)}} \partial^2_{\sigma^2} F(\sigma^2,r) + \frac{\partial r}{\partial m_i^{(a)}} \partial^2_{\sigma^2,r} F(\sigma^2,r)\Big], & \\
     \frac{\partial \omega_\mu}{\partial m_i^{(a)}} - \frac{\partial b}{\partial m_i^{(a)}} g_\mu - b\frac{\partial g_\mu}{\partial m_i^{(a)}} = \frac{\Phi_{\mu i}}{\sqrt{n}} \bbe_a,& \\ 
     \frac{\partial b}{\partial m_i^{(a)}}  = - \frac{2}{\alpha} \Big[ \frac{\partial \sigma^2}{\partial m_i^{(a)}} \partial^2_{\sigma^2,r} F(\sigma^2,r) + \frac{\partial r}{\partial m_i^{(a)}} \partial^2_{r} F(\sigma^2,r)\Big].&
\end{subnumcases}
Here we denoted $\bbe_a = 1$ if $\bbK = \bbR$, and $(\bbe_a)_b = \delta_{ab}$ if $\bbK = \bbC$.  
In particular, taken at the trivial fixed point, these equations can be greatly simplified, using the value of the parameters at this point, the symmetries of the channel and prior, and the development of the $F$ function, cf eq.~\eqref{eq:F_small_r}:
\begin{subnumcases}{\label{eq:derivative_tap_eqs_fixed_point}}
     \delta_{ij} \bbe_a =-\beta \EE_{P_0}\Big[x \big(x \cdot \frac{\partial \lambda_j}{\partial m_i^{(a)}}\big) \Big], & \\
     \frac{\partial \sigma^2}{\partial m_i^{(a)}} = \frac{-\beta}{2} \frac{\partial\gamma}{\partial m_i^{(a)}}(\EE_{P_0}[|x|^4] - (\EE_{P_0}[|x|^2])^2) , & \\ 
     \frac{\partial g_\mu}{\partial m_i^{(a)}} = \frac{\alpha}{\rho \langle \lambda \rangle_\nu}\frac{\partial \omega_\mu}{\partial m_i^{(a)}}  - \frac{\beta \alpha^2}{\rho^2 \langle \lambda \rangle_\nu^2}\EE_{P_\mathrm{out}(y_\mu,0,\rho \langle \lambda \rangle_\nu / \alpha)}\big[h \big(h \cdot \frac{\partial \omega_\mu}{\partial m_i^{(a)}} \big)\big] ,& \\
     \frac{\partial r}{\partial m_i^{(a)}} = \frac{2 \alpha^2}{\rho^2 \langle \lambda \rangle_\nu^2} \frac{\partial b}{\partial m_i^{(a)}} & \\
     \frac{\partial \gamma}{\partial m_i^{(a)}} = \langle \lambda \rangle_\nu \frac{\partial r}{\partial m_i^{(a)}} , & \\
     \frac{\partial \omega_\mu}{\partial m_i^{(a)}} - \frac{\rho \langle \lambda \rangle_\nu}{\alpha}\frac{\partial g_\mu}{\partial m_i^{(a)}} = \frac{\Phi_{\mu i}}{\sqrt{n}} \bbe_a,& \\ 
     \frac{\partial b}{\partial m_i^{(a)}}  = \frac{\langle \lambda \rangle_\nu}{\alpha} \frac{\partial \sigma^2}{\partial m_i^{(a)}} - \frac{\rho^2}{2 \alpha^2} [\alpha \langle \lambda^2 \rangle_\nu - (1+\alpha)\langle \lambda \rangle_\nu^2 ] \frac{\partial r}{\partial m_i^{(a)}}.&
\end{subnumcases}
We used eq.~\eqref{eq:identity_bo_differentiated} and eq.~\eqref{eq:bayes_optimal_gout} (from the derivation of $\bM^\mathrm{(LAMP)}$) to simplify the equation involving the derivative of $r$.
One can already notice the very interesting fact that the variance scalar parameters and the vector parameters are decoupled ! Moreover, it is easy to see that the equations on the variance parameters can be closed to:
\begin{align*}
    \frac{\partial \sigma^2}{\partial m_i^{(a)}} &= - \frac{\rho^2 (1+\alpha) (\langle \lambda^2 \rangle_\nu - \langle \lambda \rangle_\nu^2)}{\beta \langle \lambda \rangle_\nu^2 \mathrm{Var}_{P_0}[|X|^2]} \frac{\partial \sigma^2}{\partial m_i^{(a)}}. 
\end{align*}
This equation is of the type $\partial_{m_i^{(a)}} \sigma^2 = - t \partial_{m_i^{(a)}} \sigma^2$, with $t > 0$, and thus we have 
\begin{align}\label{eq:derivative_variances_fixed_point}
    \frac{\partial \sigma^2}{\partial m_i^{(a)}} &= \frac{\partial \gamma}{\partial m_i^{(a)}} = \frac{\partial r}{\partial m_i^{(a)}} = \frac{\partial b}{\partial m_i^{(a)}} = 0.
\end{align}
Moreover, from eq.~\eqref{eq:derivative_tap_eqs_fixed_point}, we can obtain as well the derivatives of the vector parameters at the trivial fixed point:
\begin{subnumcases}{}
    \frac{\partial \lambda_j}{\partial m_i^{(a)}} = - \frac{\delta_{ij}}{\rho} \bbe_a, & \\
    \frac{\partial g_\mu}{\partial m_i^{(a)}} = \frac{\alpha}{\rho \langle \lambda \rangle_\nu} \Big[1 - \frac{\alpha}{\rho \langle \lambda \rangle_\nu} \EE_{P_\mathrm{out}(y_\mu,0,\rho \langle \lambda \rangle_\nu / \alpha)}[|h|^2]\Big]\frac{\partial 
    \omega_\mu}{\partial m_i^{(a)}} \nonumber \\
     \hspace{1cm}= - \partial_\omega g_\mathrm{out}(y_\mu,0,\rho \langle \lambda \rangle_\nu / \alpha) \frac{\partial \omega_\mu}{\partial m_i^{(a)}}, & \\
     \frac{\partial \omega_\mu}{\partial m_i^{(a)}} = \frac{\rho \langle \lambda \rangle_\nu}{\alpha}\frac{\partial g_\mu}{\partial m_i^{(a)}} + \frac{\Phi_{\mu i}}{\sqrt{n}} \bbe_a.&
\end{subnumcases}
These equations can easily be solved as:
\begin{subnumcases}{\label{eq:derivatives_vector_parameters}}
    \frac{\partial \lambda_j}{\partial m_i^{(a)}} = - \frac{\delta_{ij}}{\rho} \bbe_a, & \\
    \frac{\partial \omega_\mu}{\partial m_i^{(a)}} = \frac{\Phi_{\mu i}}{\sqrt{n}} \frac{1}{1 + \frac{\rho \langle \lambda \rangle_\nu}{\alpha} \partial_\omega g_\mathrm{out}(y_\mu,0,\rho \langle \lambda \rangle_\nu / \alpha)} \bbe_a, &\\
    \frac{\partial g_\mu}{\partial m_i^{(a)}} = - \frac{\Phi_{\mu i}}{\sqrt{n}} \frac{\partial_\omega g_\mathrm{out}(y_\mu,0,\rho \langle \lambda \rangle_\nu / \alpha)}{1 + \frac{\rho \langle \lambda \rangle_\nu}{\alpha} \partial_\omega g_\mathrm{out}(y_\mu,0,\rho \langle \lambda \rangle_\nu / \alpha)} \bbe_a. &
\end{subnumcases}

\subsection{The expansion of the free entropy} 

We start from eq.~\eqref{eq:TAP_free_entropy}:
\begin{align*}
   f_\mathrm{TAP}&(\bmm) = \frac{\beta}{n} \sum_{i=1}^n \lambda_i \cdot m_i  + \frac{\beta \gamma}{2n}\big(n \sigma^2 + \sum_{i=1}^n |m_i|^2\big) + \frac{\alpha \beta}{m} \sum_{\mu=1}^m \omega_\mu \cdot g_\mu \nonumber \\ 
   &-\frac{\beta b}{2 n} \big(\sum_{\mu=1}^m |g_\mu|^2 -\alpha n r\big)  + \frac{1}{n} \sum_{i=1}^n \ln \int_\bbK P_0(\mathrm{d}x) e^{-\frac{\beta \gamma}{2} |x|^2- \beta \lambda_i \cdot x}  \\ 
   &+ \frac{\alpha}{m} \sum_{\mu=1}^m \ln \int_\bbK \frac{\mathrm{d}h}{\big(\frac{2 \pi b}{\beta}\big)^{\beta/2}} P_\mathrm{out}(y_\mu|h) e^{-\frac{\beta|h-\omega_\mu|^2}{2 b}} - \frac{\beta}{n} \sum_{i=1}^n \sum_{\mu=1}^m g_\mu \cdot \big(\frac{\Phi_{\mu i}}{\sqrt{n}}  m_i\big) + \beta F(\sigma^2,r). \nonumber
\end{align*}
At the trivial fixed point, we obtain by differentiating this expression twice (using the form of the trivial fixed point and eq.~\eqref{eq:derivative_variances_fixed_point}):
\begin{align*}
   &\frac{\partial^2 f_\mathrm{TAP}}{\partial m_i^{(a)} \partial m_j^{(b)}} = \\
   & \frac{2\beta}{n} \delta_{ij} \delta_{ab} \frac{\partial \lambda_{i}^{(a)}}{\partial m_i^{(a)}} + \frac{\alpha \beta}{m} \sum_{\mu=1}^m \Big[\frac{\partial \omega_\mu}{\partial m_i^{(a)}} \cdot \frac{\partial g_\mu}{\partial m_j^{(b)}} + \frac{\partial \omega_\mu}{\partial m_j^{(b)}} \cdot \frac{\partial g_\mu}{\partial m_i^{(a)}}\Big] + \beta \rho \delta_{ij} \delta_{ab} \Big(\frac{\partial \lambda_i^{(a)}}{\partial m_i^{(a)}}\Big)^2 \\
   &- \frac{\beta \rho \langle \lambda \rangle_\nu}{m} \sum_{\mu=1}^m \frac{\partial g_\mu}{\partial m_i^{(a)}} \cdot \frac{\partial g_\mu}{\partial m_j^{(b)}} 
   - \frac{\beta}{n} \sum_{\mu=1}^m \Big\{\frac{1}{\sqrt{n}} \frac{\partial (\overline{\Phi_{\mu i}} g_\mu)^{(a)}}{\partial m_j^{(b)}} + \frac{1}{\sqrt{n}} \frac{(\overline{\Phi_{\mu j}}\partial g_\mu)^{(b)}}{\partial m_i^{(a)}}\Big\}  \\
   &+ \frac{\alpha}{m} \sum_{\mu=1}^m \Big[-\frac{\beta \alpha}{2 \rho \langle \lambda \rangle_\nu} \frac{\partial^2 b}{\partial m_i^{(a)} \partial m_j^{(b)}} + \frac{\beta \alpha^2}{2 \rho^2 \langle \lambda \rangle_\nu^2} \frac{\partial^2 b}{\partial m_i^{(a)} \partial m_j^{(b)}} \EE_{P_\mathrm{out}(y_\mu,0,\rho \langle \lambda \rangle_\nu / \alpha)}[|h|^2]\Big] \\ 
   &+ \frac{\beta \alpha^2}{m \rho \langle \lambda \rangle_\nu} \sum_{\mu=1}^m \Big(\frac{\partial \omega_\mu}{\partial m_i^{(a)}}\Big)\cdot  \Big(\frac{\partial \omega_\mu}{\partial m_j^{(b)}}\Big) \Big\{\frac{\alpha}{\rho \langle \lambda \rangle _\nu }\EE_{P_\mathrm{out}(y_\mu,0,\rho \langle \lambda \rangle_\nu / \alpha)}[|h|^2] - 1 \Big\}.
\end{align*}
We then use eq.~\eqref{eq:derivatives_vector_parameters} and eq.~\eqref{eq:bayes_optimal_gout}, to simplify slightly the result:
\begin{align*}
   \frac{n}{\beta}\frac{\partial^2 f_\mathrm{TAP}}{\partial m_i^{(a)} \partial m_j^{(b)}} &= \frac{-1}{ \rho} \delta_{ij} \delta_{ab} + \sum_{\mu=1}^m \Big[\frac{\partial \omega_\mu}{\partial m_i^{(a)}} \cdot \frac{\partial g_\mu}{\partial m_j^{(b)}} + \frac{\partial \omega_\mu}{\partial m_j^{(b)}} \cdot \frac{\partial g_\mu}{\partial m_i^{(a)}}\Big] \\
   & - \frac{\rho \langle \lambda \rangle_\nu}{\alpha} \sum_{\mu=1}^m \frac{\partial g_\mu}{\partial m_i^{(a)}} \cdot \frac{\partial g_\mu}{\partial m_j^{(b)}} 
   - \sum_{\mu=1}^m \Big\{\frac{1}{\sqrt{n}} \frac{\partial (\overline{\Phi_{\mu i}} g_\mu)^{(a)}}{\partial m_j^{(b)}} + \frac{1}{\sqrt{n}} \frac{(\overline{\Phi_{\mu j}}\partial g_\mu)^{(b)}}{\partial m_i^{(a)}}\Big\} \\
   &+ \sum_{\mu=1}^m \Big(\frac{\partial \omega_\mu}{\partial m_i^{(a)}}\Big)\cdot  \Big(\frac{\partial \omega_\mu}{\partial m_j^{(b)}}\Big) \partial_\omega g_\mathrm{out}(y_\mu,0,\rho \langle \lambda \rangle_\nu / \alpha).
\end{align*}
We also used eq.~\eqref{eq:dwgout_trivial_fixed_point} to make $\partial_\omega g_\mathrm{out}$ appear in the last term.
As is clear from this last equation and eq.~\eqref{eq:derivatives_vector_parameters}, the dependency on $a,b$ of the result will fully be determined by the quantity $(\Phi_{\mu i} \bbe_a) \cdot (\Phi_{\mu j} \bbe_b)$. For $\beta = 1$, this is simply equal to $\Phi_{\mu i} \Phi_{\mu j}$. For $\beta = 2$, this can be represented as a $2 \times 2$ matrix:
\begin{align*}
 \big\{ (\Phi_{\mu i} \bbe_a) \cdot (\Phi_{\mu j} \bbe_b) \big\}_{a,b=1,2} = 
 \begin{pmatrix}
 \mathrm{Re}[\overline{\Phi_{\mu i}}\Phi_{\mu j}] & -\mathrm{Im}[\overline{\Phi_{\mu i}}\Phi_{\mu j}] \\
 \mathrm{Im}[\overline{\Phi_{\mu i}}\Phi_{\mu j}] & \mathrm{Re}[\overline{\Phi_{\mu i}}\Phi_{\mu j}] \end{pmatrix}.
\end{align*}
This is just the usual matrix representation of the complex number $\overline{\Phi_{\mu i}} \Phi_{\mu j}$. Following this representation, we can formally write $n\frac{\partial^2 f_\mathrm{TAP}}{\partial m_i \partial m_j}$ as an element of $\bbK$ ! This yields:
\begin{align*}
   \frac{n}{\beta}\frac{\partial^2 f_\mathrm{TAP}}{\partial m_i \partial m_j} &= \frac{-1}{ \rho} \delta_{ij} +  \sum_{\mu =1}^m \frac{\overline{\Phi_{\mu i}}\Phi_{\mu j}}{n} \frac{\partial_\omega g_\mathrm{out}(y_\mu,0,\rho \langle \lambda \rangle_\nu / \alpha)}{1 + \frac{\rho \langle \lambda \rangle_\nu}{\alpha}\partial_\omega g_\mathrm{out}(y_\mu,0,\rho \langle \lambda \rangle_\nu / \alpha)}.
\end{align*}
\section{Additional numerical experiments}\label{sec:app_numerics}

\subsection{The transition in the spectra}
\begin{figure}[ht]
     \centering
      \begin{subfigure}[b]{0.48\textwidth}
     and the signal are \centering
         \includegraphics[width=\textwidth]{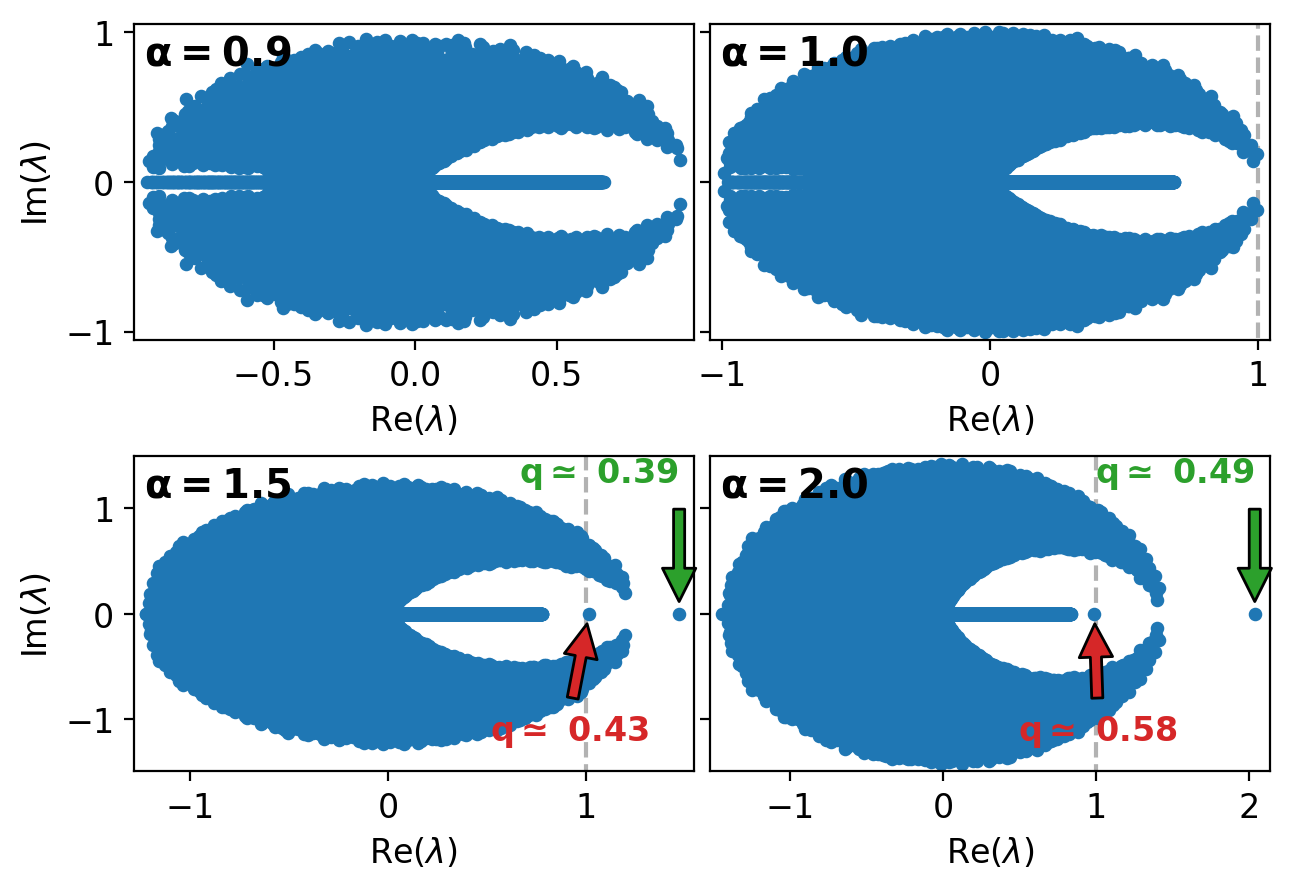}
     \caption{$\bM^{\mathrm{(LAMP)}}$\label{subfig:lamp_noiseless}}
     \end{subfigure}
      \begin{subfigure}[b]{0.48\textwidth}
        \centering
         \includegraphics[width=\textwidth]{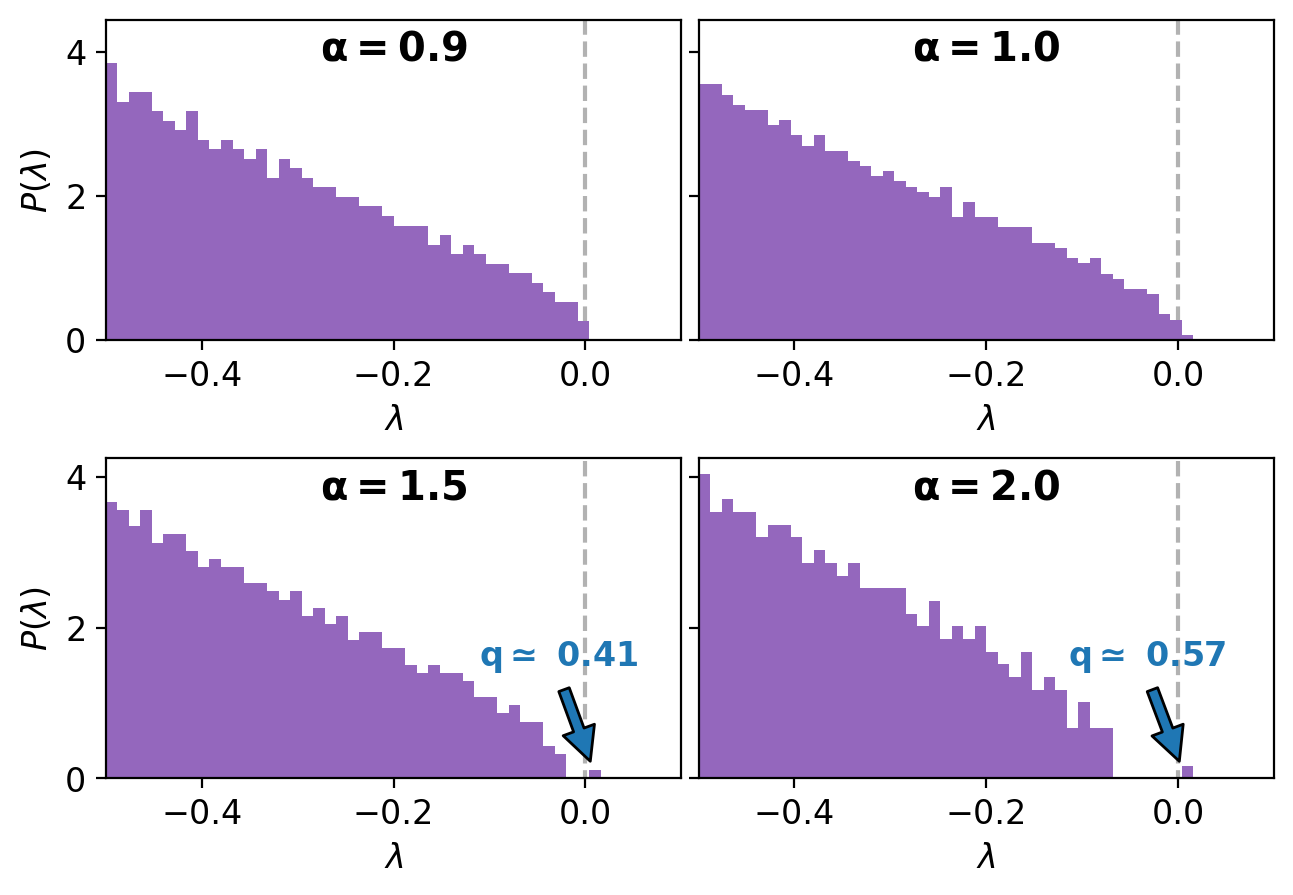}
     \caption{$\bM^{\mathrm{(TAP)}}$\label{subfig:tap_noiseless}}
     \end{subfigure}
        \caption{Transition in the spectrum of $\bM^{\mathrm{(LAMP)}}$ and $\bM^{\mathrm{(TAP)}}$ for a complex Gaussian $\bPhi$ and a noiseless phase retrieval channel. For $\alpha > \alpha_\mathrm{WR,Algo} = 1$, we indicate the approximate overlap $q$ corresponding to the relevant eigenvalues.
        \label{fig:transition_complex_gaussian_noiseless}
        }
\end{figure}
\begin{figure}[ht]
     \centering
      \begin{subfigure}[b]{0.48\textwidth}
        \centering
         \includegraphics[width=\textwidth]{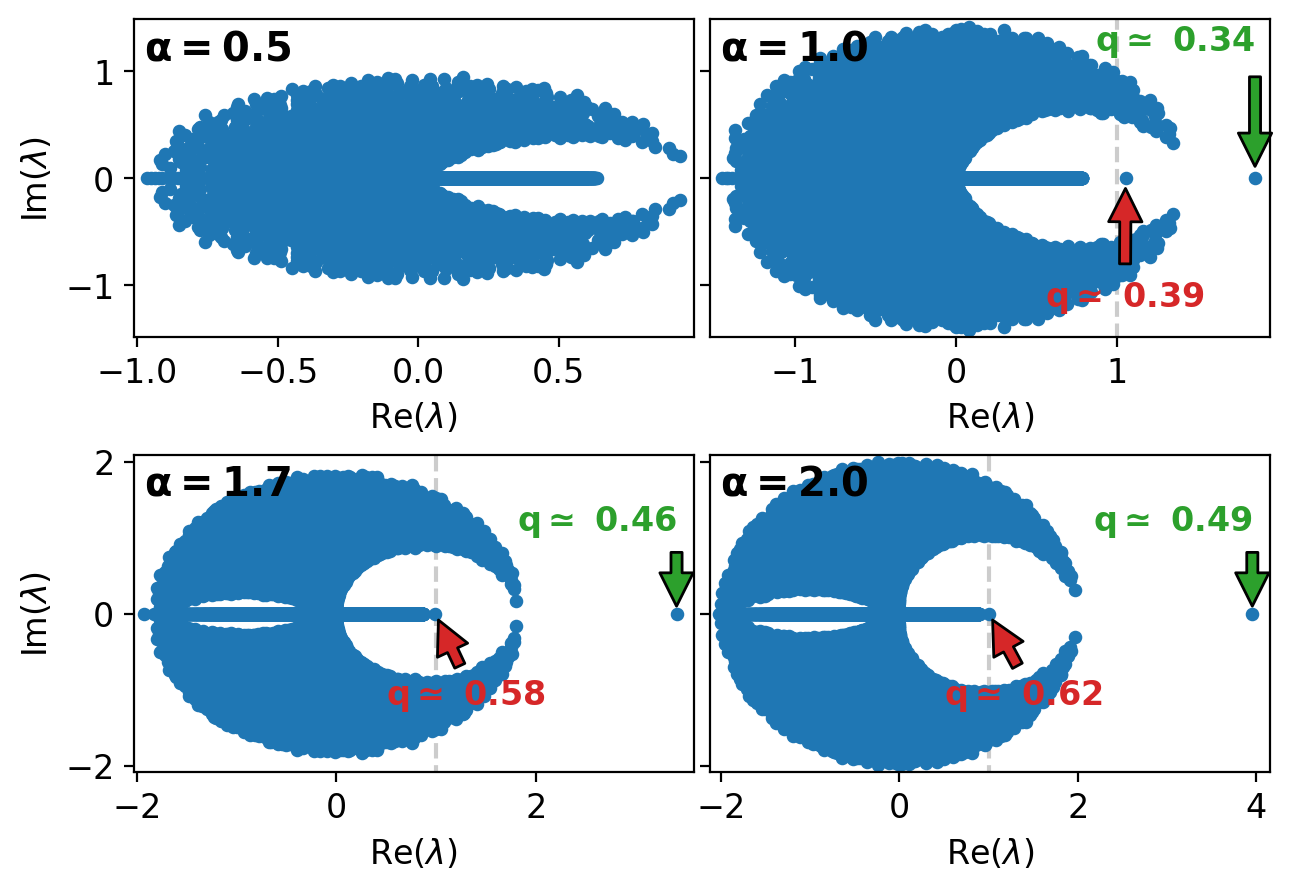}
     \caption{$\bM^{\mathrm{(LAMP)}}$\label{subfig:lamp_real_image}}
     \end{subfigure}
      \begin{subfigure}[b]{0.48\textwidth}
        \centering
         \includegraphics[width=\textwidth]{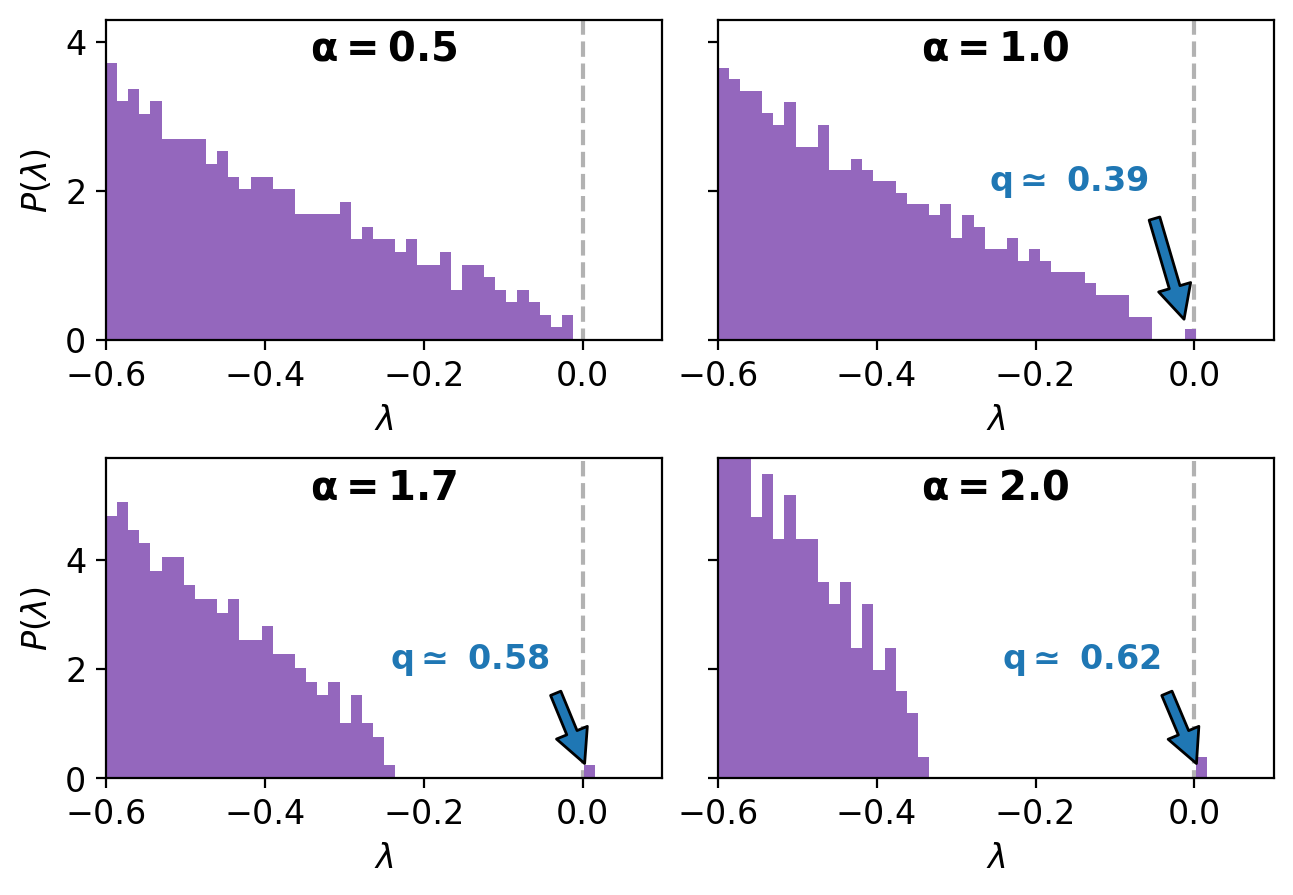}
     \caption{$\bM^{\mathrm{(TAP)}}$\label{subfig:tap_real_image}}
     \end{subfigure}
        \caption{Transition in the spectra of $\bM^{\mathrm{(LAMP)}}$ and $\bM^{\mathrm{(TAP)}}$ 
        for $\bPhi$ being the product of two complex Gaussian matrices, and a noiseless phase retrieval channel, for the recovery of a natural image.
        For $\alpha > \alpha_\mathrm{WR,Algo} = 0.5$, we indicate the approximate overlap $q$ corresponding to the relevant eigenvalues.
        \label{fig:transition_real_image}
        }
\end{figure}
\noindent
In this section, we present two additional numerical experiments illustrating the weak-recovery transition
in the spectra of $\bM^\mathrm{(TAP)}$ and $\bM^\mathrm{(LAMP)}$.
These figures are very similar to Fig.~\ref{fig:transition_complex_gaussian_poisson} in the main text.
Namely, in Fig.~\ref{fig:transition_complex_gaussian_noiseless}, we consider noiseless phase retrieval with a 
complex Gaussian matrix, and in Fig.~\ref{fig:transition_real_image} we consider noiseless phase retrieval with a product of two complex Gaussian matrices, 
and a real image signal, detailed in Section~\ref{subsec:real_image}.

\subsection{The performance of the spectral initialization used in gradient descent}\label{subsec:app_performance_spectral_GD}

\begin{figure}[ht]
     \centering
    \includegraphics[width=\textwidth]{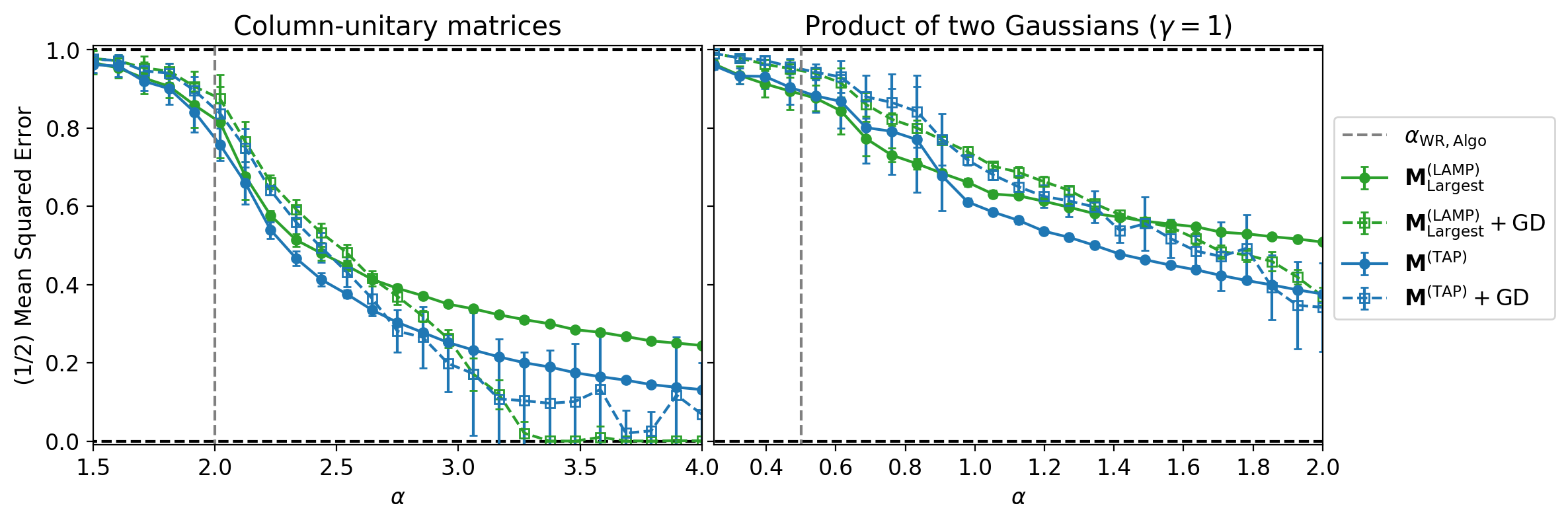} 
        \caption{
            Mean squared error achieved for the reconstruction of a real image in noiseless phase retrieval with partial DFT (left) and product of Gaussians (right) sensing matrices. 
            We compare the performance of the vanilla spectral methods and of a gradient descent procedure initialized at the spectral estimator.
            The image size was reduced from $1280 \times 820$ to $128 \times 82$. The error bars are taken over $3$ instances for each of the $3$ RGB channels.
        }
        \label{fig:mse_real_image_GD}
\end{figure}

In this section, we show the MSE achieved by a combination of our spectral methods and a gradient descent procedure for the recovery of the real image shown in Fig.~\ref{fig:comparison_TAP_with_GD}. The results are given in Fig.~\ref{fig:mse_real_image_GD}. The gradient descent procedure allows a significant improvement of the 
performance when the spectral method already achieves reasonably low error. In particular, it is able to reach perfect recovery at finite $\alpha$, which is not possible via the vanilla spectral methods.
\section{Some technicalities}\label{sec:app_technicalities}

\subsection{The linear variations of the scalar parameters}\label{subsec:app_variances_lamp}

For any quantity $r$, we write $\delta r$ its linear variation around the trivial fixed point.
One obtains the following set of equations, using the symmetry of $P_0$ and $P_\mathrm{out}$:
\begin{align}\label{eq:variation_variances}
  \begin{cases}
  \delta v_1^t = -\frac{\beta}{2} \delta \gamma_1^t\int_\bbK P_0(\mathrm{d}x) \ [|x|^4 - \rho^2], & \delta \gamma_2^t = -\frac{1}{\rho^2} \delta v_1^t - \delta \gamma_1^t,\\
  \delta c_1^t = \frac{\beta}{2} \Big[\frac{\rho^2 \langle \lambda \rangle_\nu^2}{\alpha^2}
  -\frac{1}{m}\sum\limits_{\mu=1}^m
  \frac{\int_\bbK \mathrm{d}z\ |z|^4 \ P_\mathrm{out}(y_\mu | z) \ e^{-\frac{\beta \alpha|z|^2}{2 \rho \langle \lambda \rangle_\nu} }}{\int_\bbK \mathrm{d}z\ P_\mathrm{out}(y_\mu | z) \ e^{-\frac{\beta \alpha|z|^2}{2 \rho \langle \lambda \rangle_\nu }}}
   \Big] \delta \tau_1^t, & \delta \tau_2^t = - \frac{\alpha^2}{\rho^2 \langle \lambda \rangle_\nu^2} \delta c_1^t - \delta \tau_1^t, \\
  \delta v_2^t = - \rho^2 \delta \gamma_2^t - \rho^2 \langle \lambda \rangle_\nu \delta \tau_2^t, & \delta \gamma^{t+1}_1 = - \frac{1}{\rho^2} \delta v_2^t - \delta \gamma_2^t, \\
  \delta c_2^{t} =  - \frac{\langle \lambda \rangle_\nu}{\alpha}\rho^2 \delta \gamma_2^t - \frac{\rho^2}{\alpha} \langle \lambda^2 \rangle_\nu \delta \tau_2^t, & \delta \tau_1^{t+1} = - \frac{\alpha^2}{\rho^2 \langle \lambda \rangle_\nu^2} \delta c_2^t - \delta \tau_2^t. 
  \end{cases}
\end{align}
Note that the linear variations of these scalar variance parameters do not depend on the variations of the vector parameters of Algorithm~\ref{algo:gvamp}.
Differentiating eq.~\eqref{eq:bayes_optimal_gout} with respect to $\tau_1^t$ and taking it 
at the trivial fixed point implies:
\begin{align}\label{eq:identity_bo_differentiated}
  \textstyle 
\frac{1}{m}\sum_{\mu=1}^m
\Big[\frac{\int_\bbK \mathrm{d}z\ |z|^4 \ P_\mathrm{out}(y_\mu | z) \ e^{-\frac{\beta \alpha}{2 \rho \langle \lambda \rangle_\nu} |z|^2}}{\int_\bbK \mathrm{d}z\ P_\mathrm{out}(y_\mu | z) \ e^{-\frac{\beta \alpha}{2 \rho \langle \lambda \rangle_\nu |z|^2}}} - \Big\{\frac{\int_\bbK \mathrm{d}z\ |z|^2 \ P_\mathrm{out}(y_\mu | z) \ e^{-\frac{\beta \alpha}{2 \rho \langle \lambda \rangle_\nu} |z|^2}}{\int_\bbK \mathrm{d}z\ P_\mathrm{out}(y_\mu | z) \ e^{-\frac{\beta \alpha}{2 \rho \langle \lambda \rangle_\nu |z|^2}}}\Big\}^2 \Big]&= \frac{2 \rho^2 \langle \lambda \rangle_\nu^2}{ \beta \alpha^2}.
\end{align}
Using this relation, one obtains from eq.~\eqref{eq:variation_variances} that $\delta c_1^t = - \delta \tau_1^t \rho^2 \langle \lambda \rangle_\nu^2 / \alpha^2$,
which then implies $\delta \tau_2^t = 0$. Similarly, it follows easily by the remaining
equations that all the variations in eq.~\eqref{eq:variation_variances} must be zero.

\subsection{The expansion of \texorpdfstring{$F(x,y)$ around $y = 0$}{}}\label{subsec:app_expansion_F}

We describe here the behavior of $F(x,y)$ as $x> 0$ and $y \to 0^+$.
Let us write the equations satisfied by $\zeta_x,\zeta_y$:
\begin{subnumcases}{\label{eq:saddle_point_zeta}}
   \Big \langle \frac{\zeta_y}{\zeta_x \zeta_y + \lambda} \Big \rangle_\nu = x, &\\
   \frac{\alpha-1}{\zeta_y} + \Big \langle \frac{\zeta_x}{\zeta_x \zeta_y + \lambda} \Big \rangle_\nu = \alpha y . &
\end{subnumcases}
As $y \to 0^+$, this implies necessarily that $\zeta_y \to + \infty$, and one finds easily that $\zeta_y \sim 1/y$, $\zeta_x \sim 1/x$.
We now turn to the next order variations, that we write as:
\begin{align*}
  \zeta_y &= y^{-1} + c_1 + {\cal O}(y),\\
  \zeta_x &= \frac{1}{x} + c_2 y  + {\cal O}(y^{2}).
\end{align*}
We use eq.~\eqref{eq:saddle_point_zeta} to compute $c_1 = - x \braket{\lambda}_\nu / \alpha$ and $c_2 = - \braket{\lambda}_\nu$.
We can then develop the logarithmic potential:
\begin{align*}
  \frac{1}{2} \langle \log(\zeta_x \zeta_y + \lambda) \rangle_\nu &= -\frac{1}{2} \log y - \frac{1}{2} \log x - \frac{x}{2\alpha} \braket{\lambda}_\nu y + {\cal O}(y^2).  
\end{align*}
Developing the other terms involved in $F(x,y)$ is straightforward and yields:
\begin{align}\label{eq:F_y0}
  F(x,y) &= - \frac{xy}{2} \braket{\lambda}_\nu + {\cal O}(y^2). 
\end{align}
One can push this analysis to the next order, and finds in the exact same way, from eq.~\eqref{eq:saddle_point_zeta}:
\begin{align*}
   \zeta_y &= \frac{1}{y} - \frac{x \braket{\lambda}_\nu}{\alpha} + \frac{x^2}{\alpha^2} \left[\alpha \braket{\lambda^2}_\nu - (1+\alpha) \braket{\lambda}_\nu^2\right]y + {\cal O}(y^2), \\ 
   \zeta_x &= \frac{1}{x} - \braket{\lambda}_\nu y + \frac{x}{\alpha} \left[\alpha \braket{\lambda^2}_\nu - (1+\alpha) \braket{\lambda}_\nu^2\right] y^2 + {\cal O}(y^3).
\end{align*}
This yields for $F(x,y)$:
\begin{align*}
  F(x,y) = - \frac{\braket{\lambda}_\nu}{2} x y + \frac{x^2}{4 \alpha}  \left[\alpha \braket{\lambda^2}_\nu - (1+\alpha) \braket{\lambda}_\nu^2\right] y^2 + {\cal O}(y^3),
\end{align*}
which concludes our analysis.

\subsection{Proof of Proposition~\ref{prop:relating_lamp_tap}}\label{subsec_app:equivalence_lamp_tap}

Let us recall the two spectral methods $\bM^{\mathrm{(TAP)}}$, $\bM^{\mathrm{(LAMP)}}$. Without loss of generality, we assume $\langle \lambda \rangle_\nu = \alpha$. Recall that 
we defined $z_\mu \equiv \partial_\omega g_\mathrm{out}(y_\mu, 0, \rho)$. We let $\bZ = \mathrm{Diag}(z_\mu)$.
We can thus write:
\begin{subnumcases}{}
    \bM^{(\text{LAMP})} = \rho \Big(\frac{\bPhi \bPhi^\dagger}{n} - \mathds{1}_m \Big) \bZ , & \\
    \bM^{\mathrm{(TAP)}} = - \frac{1}{\rho} \mathds{1}_n + \frac{1}{n} \bPhi^\dagger \frac{\bZ}{\mathds{1}_m + \rho \bZ} \bPhi . &
\end{subnumcases}
We start by the first claim.
By definition of $(\lambda_\mathrm{LAMP},\bv)$, we have 
\begin{align}\label{eq:def_ev_lamp}
  \rho \frac{\bPhi \bPhi^\dagger}{n} \bZ \bv = (\rho \bZ + \lambda_\mathrm{LAMP}) \bv. 
\end{align}
Since we assumed $\lambda_\mathrm{LAMP} + \rho z_\mu \neq 0$ for all $\mu$, this implies that $\bPhi^\dagger \bZ \bv \neq 0$, and we thus let 
\begin{align*}
  \bxhat \equiv \frac{\bPhi^\dagger \bZ \bv}{\lVert \bPhi^\dagger \bZ \bv \rVert} \sqrt{n}.
\end{align*}
Multiplying eq.~\eqref{eq:def_ev_lamp} by $\bPhi^\dagger \bZ (\lambda_\mathrm{LAMP} + \rho \bZ)^{-1}$ on both sides, we directly reach the sought result:
\begin{align*}
 \Big\{\frac{1}{n}\bPhi^\dagger \frac{\rho \bZ}{\lambda_\mathrm{LAMP} + \rho \bZ} \bPhi \Big\} \bxhat&= \bxhat.
\end{align*}
We move on to the second claim.
Let $\bx \in \bbK^n$ be an eigenvector of $\bM^\mathrm{(TAP)}$ with norm $\lVert\bx\rVert^2 = n$, with associated eigenvalue $\lambda_\mathrm{TAP}$.
We let:
\begin{align*}
  \bu \equiv \frac{\mathds{1}_m}{\mathds{1}_m + \rho \bZ} \frac{\bPhi}{\sqrt{n}} \bx.
\end{align*}
And we can then easily compute:
\begin{align}\label{eq:MLAMP_z}
  \nonumber
  \bM^\mathrm{(LAMP)} \bu &= \rho \Big(\frac{\bPhi \bPhi^\dagger}{n} - \mathds{1}_m \Big) \frac{\bZ}{\mathds{1}_m + \rho \bZ} \frac{\bPhi}{\sqrt{n}} \bx, \\
  \nonumber
  &= \frac{\rho \bPhi}{\sqrt{n}} \Big[\bM^\mathrm{(TAP)} + \frac{1}{\rho} \mathds{1}_n \Big] \bx - \rho \bZ \bu, \\
  \nonumber
  &= \rho \lambda_\mathrm{TAP} \frac{\bPhi}{\sqrt{n}} \bx + \frac{\bPhi}{\sqrt{n}} \bx - \rho \bZ \bu, \\
  \bM^\mathrm{(LAMP)} \bu &= \bu + \rho \lambda_\mathrm{TAP} (\mathds{1}_m + \rho \bZ) \bu.
\end{align}
At $\alpha = \alpha_{\mathrm{WR,Algo}}$, the largest eigenvalue of $\bM^\mathrm{(TAP)}$ concentrates on $0$, which corresponds to the onset 
of marginal instability of the trivial local maximum. As one can see from eq.~\eqref{eq:MLAMP_z}, this implies that $\bM^\mathrm{(LAMP)}$ also possesses an eigenvalue 
equal to $1$ at $\alpha = \alpha_\mathrm{WR,Algo}$, indicating marginal instability as well.
To put it shortly, \emph{the two spectral methods have the same weak recovery threshold}. Moreover, eq.~\eqref{eq:MLAMP_z} implies that for any $\alpha \geq \alpha_{\mathrm{WR,Algo}}$, if $\bM^\mathrm{(TAP)}$ has en eigenvalue that concentrates on $0$ as $n \to \infty$, 
then $\bM^\mathrm{(LAMP)}$ has a corresponding eigenvalue concentrating on $1$, and \emph{with the same performance}.
Indeed, as described in eq.~\eqref{eq:correspondance_LAMP_TAP_estimators}, the estimator associated to $\bM^\mathrm{(LAMP)}$ will be given by:
\begin{align*}
  \bxhat_\mathrm{LAMP} &\propto \frac{\bPhi^\dagger}{\sqrt{n}} \bZ \bu
  = \frac{\bPhi^\dagger}{\sqrt{n}} \frac{\bZ}{\mathds{1}_m + \rho \bZ} \frac{\bPhi}{\sqrt{n}} \bxhat_\mathrm{TAP}, 
\end{align*}
in which $\bxhat$ is an eigenvector of $\bM^\mathrm{(TAP)}$ with eigenvalue $0$. Therefore, we reach that $\bxhat_\mathrm{LAMP} \propto \bxhat_\mathrm{TAP}$, and these two vectors are thus equal 
as they are both normalized.

\end{document}